\documentclass[reqno,12pt]{article}
\usepackage{amsmath}
\usepackage{bm}
\usepackage{a4wide,amssymb}
\usepackage{graphicx,xcolor}
\usepackage{epstopdf}
\usepackage{bbm}
\usepackage{mathrsfs}

\usepackage[titletoc,toc,title]{appendix}

\usepackage[refpage]{nomencl}
\usepackage{etoolbox}

 \newcommand{\nn}{\nonumber}
\newcommand{\R}{{\mathbb R}}

\renewcommand{\a}{\alpha}

\newcommand{\g}{\gamma}

\def\s{\sigma}
\def\r{{\rho}}

\def\L{{\Lambda}}

\def\be{\begin{equation}}
\def\ee{\end{equation}}
\def\bea{\begin{eqnarray}}
\def\eea{\end{eqnarray}}

\def\nn{\nonumber}

\def\o{\omega}
\def\b{\beta}

\usepackage{multicol}
\makenomenclature
\usepackage{xpatch}
\makeatletter
\xapptocmd\thenomenclature{\let\@item\nomencl@item\def\nomencl@width{0pt}}{}{}
\let\nomencl@item\@item
\xpretocmd\nomencl@item{\nomencl@measure{#1}}{}{}
\def\nomencl@measure#1{%
  \sbox0{#1}%
  \ifdim\wd0>\nomencl@width\relax
    \edef\nomencl@width{\the\wd0}%
  \fi
}

\xpatchcmd\thenomenclature
  {\section*{\nomname}}
  {\begin{multicols}{2}[\section*{\nomname}]}
  {}{}
\xpatchcmd\thenomenclature
  {\chapter*{\nomname}}
  {\begin{multicols}{2}[\chapter*{\nomname}]}
  {}{}

\xapptocmd\endthenomenclature{%
  \immediate\write\@mainaux{\global\nomlabelwidth\nomencl@width\relax}%
  \end{multicols}
}{}{}
\makeatother

\begin{document}

\begin{titlepage}

\begin{center} 
{\bf \Large{Inhomogeneities in Boltzmann--SIR models}}

\vspace{1.5cm}
{\large A. Ciallella$^{1}$, M. Pulvirenti$^{2}$ and S. Simonella$^{3}$  }
\vspace{0.5cm}

{$1.$ \scshape {\small Dipartimento di Ingegneria Civile, Edile -- Architettura e
             Ambientale, and 
             \\International Research Center M\&MOCS, 
             Universit\'a dell'Aquila,\\
             via Giovanni Gronchi 18, 67100, L'Aquila, Italy.}
\smallskip

$2.$ {\small Dipartimento di Matematica, Universit\`a di Roma La Sapienza\\ 
Piazzale Aldo Moro 5, 00185 Rome -- Italy,  and \\
 International Research Center M\&MOCS, Universit\`a dell'Aquila, \\ 
 Piazzale Ernesto Pontieri 1, Monteluco di Roio, 67100 L'Aquila -- Italy.}  
 \smallskip

$3.$ {\small UMPA UMR 5669 CNRS, ENS de Lyon \\ 46 all\'{e}e d'Italie,
69364 Lyon Cedex 07 -- France}}

\end{center}

\vspace{1.5cm}
\noindent
{\bf Abstract.} We investigate, by means of numerical simulations, the qualitative properties of a Boltzmann equation for three species of particles introduced in previous work, 
capturing some features of epidemic spread.

\thispagestyle{empty}


\bigskip
\bigskip

\newpage



\end{titlepage}


\section{Introduction} \label{sec:1}
\setcounter{equation}{0}    
\def\theequation{1.\arabic{equation}}

In a recent contribution \cite{PS_SIR}, we presented a kinetic model for mixtures of three species of particles, or ``agents'', labelled $S$, $I$, $R$. This stays for
susceptible, infected and recovered, as inspired from the basic SIR system in epidemiology. 
In this model, together with  collisions and transport, a reaction
 $$
  S+I \to I+I
 $$
 takes place with constant rate $\b\in(0,1]$, upon contact of a particle of type $S$ and a particle of type $I$. No other reactions occur, but particles of type $I$ decay as $I\to R$ with a constant rate $\g >0$. 
The one-particle distribution functions depend on time $t$, position $x$ and velocity $v$:
\bea
& f_S=f_S(t,x,v)\nn\\ 
& f_I=f_I(t,x,v)\nn\\
& f_ R=f_R(t,x,v)\nn \\
&f:=f_S+f_I+f_R \nn
\eea 
where $f$ is the total density.
The microscopic model is further based on a few elementary features:

\smallskip
\noindent $*$ the interactions are binary, and localized;

\smallskip
\noindent $*$ the number of interactions per unit time is expected to be finite;

\smallskip
\noindent $*$ the qualitative behaviour is independent of the number of particles  $N$, provided that this is large in a suitable scaling limit;

\smallskip
\noindent $*$ a statistical description is appropriate.

\smallskip
\noindent
This leads to a Boltzmann equation which reads (in two dimensions):
\begin{equation}
\begin{cases}
  \left(\partial_t + v \cdot \nabla_x\right) f_S= Q(f_S,f_S) + Q(f_S,f_R)+(1-\b)Q(f_S,f_I)-\b Q_-(f_S,f_I)  \\
    \left(\partial_t + v \cdot \nabla_x\right) f_I = Q(f_I,f_I) + Q(f_I,f_R)+Q(f_I,f_S)+\b Q_+(f_S,f_I)-\g f_I\\
     \left(\partial_t + v \cdot \nabla_x\right) f_R = Q(f_R,f)+\g f_I\\
\end{cases} \label{eq:SIR_Boltzmann}
\end{equation}
where 
\bea
&& Q= Q_+ - Q_- \;,\nn\\
&& Q_+(f,g)(v) := \int_{\R^2} \int_{{\mathbb S}} B(\o; v-v_*)f(v')g(v'_*) \, d\o \, dv_*\;, \nn\\
&& Q_-(f,g)(v) := f(v)\int_{\R^2} \int_{{\mathbb S}} B(\o; v-v_*)g(v_*) \,  d\o \, dv_*\;, \nn
\eea
$B$ is a given interaction kernel (see the next section for specific choices) and typically
\be
\begin{cases}
 v'=v-\o (v-v_*)\cdot \o\\
 v'_*=v_*+ \o (v-v_*)\cdot \o
\end{cases}\;\label{coll}
\ee
are the outgoing velocities for a collision, preserving momentum and energy.
In particular, $f$ is governed by the standard Boltzmann equation:
\be
\label{trueB} 
\left(\partial_t + v \cdot \nabla_x\right) f = Q(f,f)\;.
\ee


%

The formal link with the classical  theory of epidemics is obtained for spatially homogeneous distributions
(no dependence on $x$), looking at averaged fractions of agents $A(t):=\int  f_A(t,v) dv $  in the species 
$A \in \{ S,I,R \}$. Performing the integral with respect to $v$ of Eq.s\,\eqref{eq:SIR_Boltzmann} and using that $\int Q_+=\int Q_-$, we find
\be \label{eq:SIR''}
\begin{cases}
  \dot S = -\b \,Q_- (f_S,f_I)  \nn\\
    \dot I = \b\, Q_- (f_S,f_I) -\g I \nn\\
     \dot R = \g I \nn\\
\end{cases}\;.
\ee
Such a set of equations is not closed but, when dealing with ``Maxwellian molecules'' defined by the requirement that $ \int B\, d\o =1$, one gets
$Q_- (f_S,f_I) = S(t)I(t)$ and therefore
\be \label{eq:SIR}
\begin{cases}
  \dot S = -\b IS  \\
    \dot I = \b IS -\g I \\
     \dot R = \g I \\
\end{cases}\;;
\ee
namely the simplest SIR model equations \cite{KMK}. The latter have been considerably used and extended, for almost a century; see e.g.\,\cite{BC-C01} for a recent overview of mathematical epidemiology, or \cite{Quart21} for a case in the huge amount of studies on the current COVID-19 pandemic. 

A stochastic $N$-particle system can be constructed, with distribution functions converging to the solution of \eqref{eq:SIR_Boltzmann} as $N \to \infty$ (\cite{PS_SIR}) and corresponding, numerically,
to the DSMC (direct Simulation Monte Carlo) method. In the present paper, we adopt 
the Boltzmann-SIR equations \eqref{eq:SIR_Boltzmann} as toy model, and the underlying particle
system as a tool to study the qualitative behaviour. This allows to reinterpret 
some features of SIR type models, in terms of 
spatial inhomogeneities.  

As in \cite{PS_SIR} we stress that we do not pretend the kinetic model to provide any realistic prediction in epidemiology,
as of course real agents do not interact as elastically colliding particles in a rarefied gas.
Realistic interactions are obviously difficult to be described in mathematical terms. Individual strategies might play a critical role and, in essence, the interactions might be not even necessarily binary
(e.g.\,a single agent infecting many susceptible agents almost simultaneously).
Motivated by the simple connection with \eqref{eq:SIR}, we are rather interested in capturing behaviour which has only little dependence on the details of the microscopic interaction. 

More precisely, we perform numerical simulations of system \eqref{eq:SIR_Boltzmann}, with the following plan. In Section \ref{sec:2}, we consider several kernels $B$ and verify that: (i) the macroscopic evolution for $(S(t),I(t),R(t))$ is rather insensitive to the choice of the cross-section; 
(ii) the evolution can be significantly sensitive to spatial non-uniformity of labels, even when $f$ has reached global equilibrium. 
In Section \ref{sec:3}, we perturb the model by external actions mimicking, roughly, meeting points with (airport, travel stations) or without (supermarket) injection of agents.  We observe how the local concentration of densities enhances the transient of $I(t)$, and identify regimes for the external flows producing nontrivial asymptotic values, and possibly recurrent waves.

\section{The free model} \label{sec:2}
\setcounter{equation}{0}    
\def\theequation{2.\arabic{equation}}

This section is devoted to the basic properties of Eq.\,\eqref{eq:SIR_Boltzmann}, referred to as 
``free model'' (model without external actions), which we recall (in more compact form):
\begin{equation}
\begin{cases}
  \left(\partial_t + v \cdot \nabla_x\right) f_S= Q(f_S,f)-\b Q_+(f_S,f_I)  \\
    \left(\partial_t + v \cdot \nabla_x\right) f_I = Q(f_I,f) +\b Q_+(f_S,f_I)-\g f_I\\
     \left(\partial_t + v \cdot \nabla_x\right) f_R = Q(f_R,f)+\g f_I.\\
\end{cases}\;. \nn\label{eq:SIR_BoltzmannC}
\end{equation}
We shall consider three different cross--sections, namely:

\bigskip
 \noindent {\it 1) Hard spheres}, as for the mechanical system of $N$ billiard balls, from which 
 the Boltzmann equation is obtained in the Boltzmann--Grad limit (see e.g.\,\cite {CIP}).
 The collision law is given by \eqref{coll} and the interaction kernel is
\be
\label{B}
B(\o; v-v_*)=\o \cdot(v-v_*)  \,\,\,\, \chi (\o \cdot(v-v_*) \geq 0)\;,
\ee
where $\chi (E)$ is the characteristic function of the event $E$.
  
  \medskip
\noindent   {\it 2) Semidiscrete model}, which is again a hard--sphere type system, but with particle velocities of modulus $1$, i.e.\,$v \in {\mathbb S}$. The collision law is
   \be
 \label{colls}
\begin{cases}
 v'=v-2\o \left(v\cdot \o\right) \\
 v'_*=v_*-2\o \left(v_*\cdot \o\right)
\end{cases}\;.\nn
\ee
That is, each particle is reflected against the line orthogonal to the versor $\o$ joining the two centers. Energy is conserved but not momenta. $B$ is still given by \eqref{B}.

\medskip
\noindent {\it 3) Maxwellian molecules}. A popular simple model for the Boltzmann equation \cite{Bo88},
for which the collision law is given by \eqref{coll} while $B$ satisfies 
$$
\int d\o\, B(\o; v-v_*) =\text{const.}\,.
$$
The right hand side is remarkably independent of the relative velocity $v-v_*$. 

\bigskip

Since the full  probability density satisfies \eqref{trueB}, in cases {\it 1} and {\it 3} if the mean--free path is small
$$ 
f(x,v) \approx \frac 1{|\L|} \,M(v)
$$
after a brief transient, where $|\L|$ is the measure of the domain $\L$ and $M$ is a Maxwellian 
velocity distribution
\be \label{eq:maxwellian}
M(v)= \,\, \frac {e^{- \frac {(v-u)^2} {2 \s^2 }}} { (2 \pi \s^2 )}\;,\quad v\in \R^2
\ee
with $u \in \R^2$ and $\s>0$ determined by the initial data.
If the distribution of the labels $S$, $I$, $R$ is also independent of $x$, then
$$ 
f_A(t,v) \approx   \frac {A(t)} {|\L|} \,M(v), \qquad A=S,I,R\;.
$$
Even if the full system is at equilibrium, the dynamics of particle labels (state of the agents) may well be active and we find
\be \label{eq:SIR'}
\begin{cases}
  \dot S = -\b m IS  \\
    \dot I = \b m IS -\g I \\
     \dot R = \g I \\
\end{cases}
\ee
where
$$
m=c \int dv \int dv_* |v-v_*|^b M(v) M(v_*)\,,\qquad c >0
$$
with $b=1$ in case {\it 1}, $b=0$ in case {\it 3}.
In case  {\it 2} one has similar behaviour, but $M$ is replaced by the uniform distribution
$$ 
f(x,v) \approx \frac 1{|\L|} \frac{1}{|{\mathbb S}|}\;.
$$

Therefore the kinetic picture plays a role for a short transient only and, for a larger scale of time, it does not say more than the standard SIR model,  if the distributions of labels are spatially homogeneous.

We recall that \eqref{eq:SIR} is almost explicitly solvable. The asymptotic distribution is found by
setting
$$
R(t)=R_0 + \g \int_0^t I(\tau) d\tau 
$$
(showing  $I(t) \to 0$ as $t \to \infty$) and
$$
\frac {d S}{dR}= - \frac \b \g S\;.
$$
Setting $ \bar A=\lim_{t \to \infty} A(t)$,  using $\bar R+ \bar S=1$
and assuming $R_0=R(0)=0$ (no recovered agents at time zero), one gets
$
\bar S=S_0 e^{- \frac \b \g \bar R}= S_0 e^{- \frac \b \g } e^{ \frac \b \g  \bar S},
$
hence
\be \label{eq:asy-S}
e^{ -\frac \b \g \bar S} \frac \b \g  \bar S= (1-I_0) \frac \b \g e^{- \frac \b \g }\;.
\nonumber
\ee
Since $\max y e^{-y}=\frac 1 e$, given $\b$ and $\g$ one can find non vanishing solutions for $\bar S$. 

In the numerical simulations of the Boltzmann-SIR model, the above asymptotic is determined, roughly,
by a ``herd immunity'' situation which is reached when susceptible agents are surrounded by a sufficiently large fraction of recovered agents, shielding them from the infected population.
%

Inspired by the fact that problems of interest are frequently non--homogeneous, we will focus now on profiles where particle labels are not uniformly distributed in space, so that the kinetic model is indeed more detailed than \eqref{eq:SIR'}.
%
%
Consider, for instance, the case of an initial distribution of infected agents concentrated in a small region. 
Even when $f$ has reached global equilibrium, the system as a whole can still be far from uniform (in space) for quite a long time
$$ 
 f_A(t,x,v) \approx \frac{ A(t,x)}{|\Lambda|}\, M(v), \qquad A=S,I,R
$$
and $(S(t),I(t),R(t))$ can be notably different from the solution of \eqref{eq:SIR'}. Such a behaviour will be discussed in the next subsection.

\subsection{Description of the simulations}\label{ssec:simul_free}
The numerical simulations are based on the  DSMC method.
The details of this method can be found for instance in Chapter 10 of \cite{CIP} or in \cite{RW05}. Here we just describe the setting.

The system consists of $N$ point particles moving in a square $\Lambda$ with side length $L=10^3$ and periodic boundary conditions. An equally spaced grid partitions the domain into identical square cells of size $\delta\times\delta$ (a total of $\left(\frac{L}{\delta}\right)^2$ cells).
$N$ is constant in time, no agent is introduced or removed from the system.
Each particle moves with constant velocity up to the next collision instant. The mean free path, i.e.\,the average distance travelled by each agent between two consecutive collisions, will be denoted by $\lambda$, the mean free time by  $\tau$.
Time is discretized and the evolution of the system is divided into a free evolution step where all particles move freely for a discrete time unit $\bar{t}$, and a collision simulation step, where pairs of particles lying in the same cell are randomly chosen to perform a binary collision.

For each simulation presented, we report the relevant parameters in the captions of the figures.
We list now the choices that are common to all simulation runs.

The particles are initially distributed uniformly in space,
while velocities are distributed uniformly on ${\mathbb S}$.
We assign to each particle a label,  $S$, $I$, or $R$, that can be distributed in both uniform or  non-uniform way, as specified in each case.
 As the energy of the system is fixed, in the case of hard spheres and Maxwellian molecules the velocity distribution quickly  converges  to a Maxwellian as \eqref{eq:maxwellian} with $u=0$ and $\sigma^2=1/2$.
  Hence $\tau\simeq\frac{\lambda}{\langle v \rangle}$ where the mean scalar velocity $\langle v \rangle$ can be explicitly calculated. 
The cells side is $\delta\simeq\frac{\lambda}{3}$ and the discrete time step is $\bar{t}\simeq\frac{\tau}{4}$.
 The number of agents $N$ is always such that, on average, at least $20$ particles lie in each cell.
The prescribed rule for the dynamics, following \eqref{eq:SIR_Boltzmann}, is that a collision of an $I$ (infected) and an $S$ (susceptible) particle produces two $I$ particles with probability $\b$, and that an $I$ particle becomes an $R$ (recovered) particle after an exponential time with rate $\g$.
The fractions at time $0$ are assumed to be $I(0)=0.005$, $S(0)=1-I(0)=0.995$, $R(0)=0$.

In the following, a few examples of numerical experiments of the system with different cross--sections are presented.
The evolution of the fractions of the three populations $S$, $I$, and $R$ is plotted for the particle system, and compared with the solution of the SIR model Eq.\,\eqref{eq:SIR'}.

We consider 
two different situations. In the first one, the initial distributions of agents are all uniform. 
In this setting, we want to test the consistency of the kinetic model with the SIR model, and check that results turn out to be independent of the choice of the cross--section.
 In the second case, the initial datum is such that all the infected agents are contained in a small disk of area $0.005\cdot L^2$. This is used to show that, even in a simple setting, the average description produced by the SIR model can lose quantitative and qualitative information related to spatial patterns.
 
\begin{figure}[ht!]
\begin{picture}(200,170)(30,150)
\put(-20,-10){
  \includegraphics[width=0.75\textwidth]{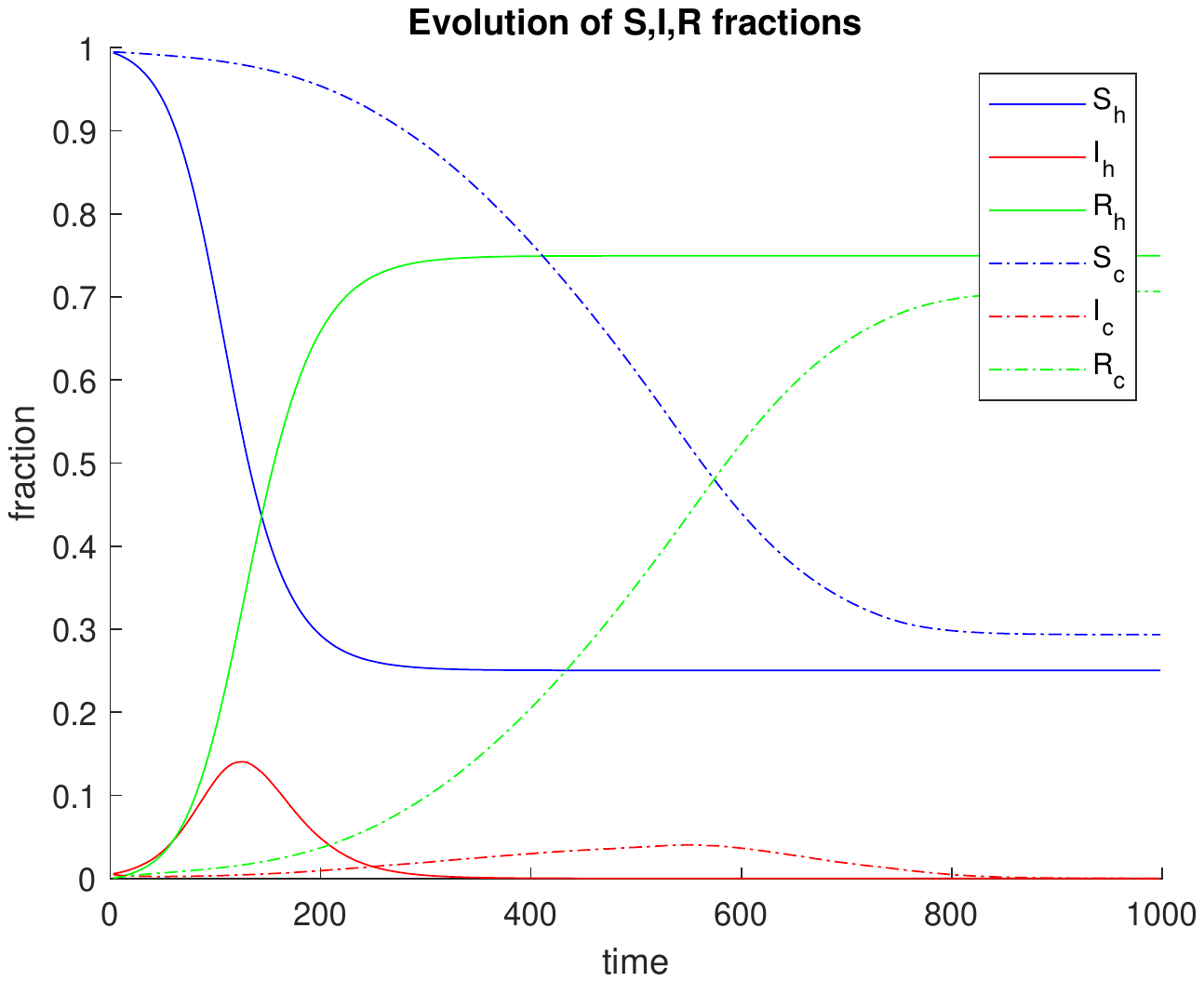}
}
\put(275,140){
  \includegraphics[width=0.38\textwidth]{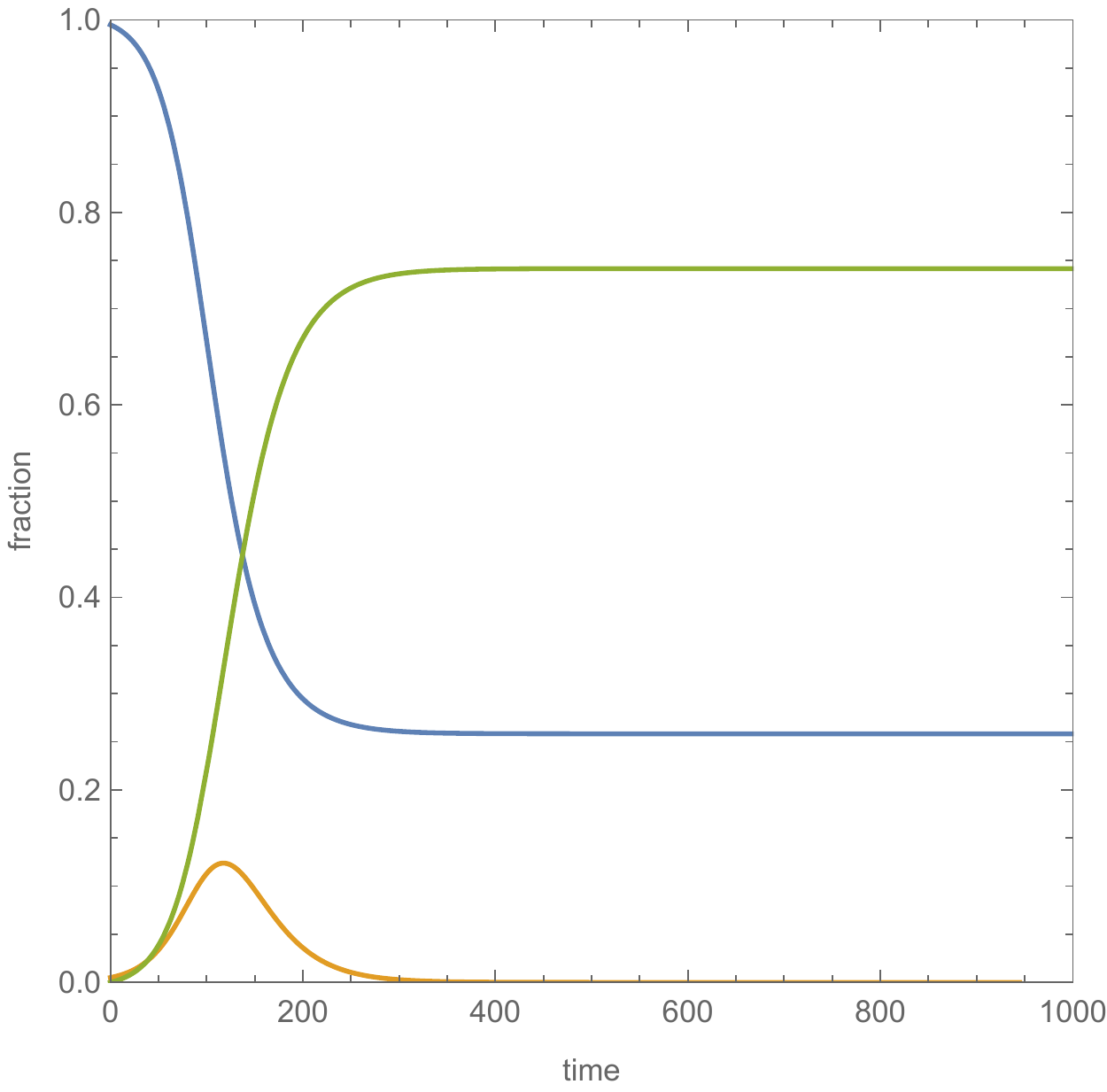}
}
\put(450,260){
  \includegraphics[width=0.06\textwidth]{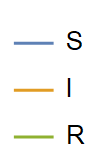}
  }
\end{picture}
\caption{
Left panel: evolution of $S$, $I$, $R$  fractions for a particle system with Maxwellian molecule cross--section simulated via DSMC. The solid and dashed lines represent, respectively, the case of homogeneous and concentrated initial population of $I$ agents. Parameters: $N=1800000$, $\b=1$, $\g=\frac{1}{20}$, $\lambda\simeq9.8$, $\tau\simeq11.05$.
Right panel: numerical solution of \eqref{eq:SIR'} with $m=\frac{1}{\tau}$, and the same $\b$ and $\g$ as in the left panel case.
}
\label{f:001}
\end{figure}

 
The case of Maxwellian molecules is reported in Fig.~\ref{f:001}. 
For uniform initial distributions, the correspondence between particle system (left panel, solid lines) and SIR (right panel) is clear.
In the left panel, we show also the case of concentrated initial distribution of infected agents (dashed lines). In this run, the mean free path is sufficiently small ($\lambda\simeq\frac{L}{100}$) to produce an apparent difference. 
%

\begin{figure}[pht!]
\begin{picture}(200,170)(30,150)
\put(-20,-10){
  \includegraphics[width=0.75\textwidth]{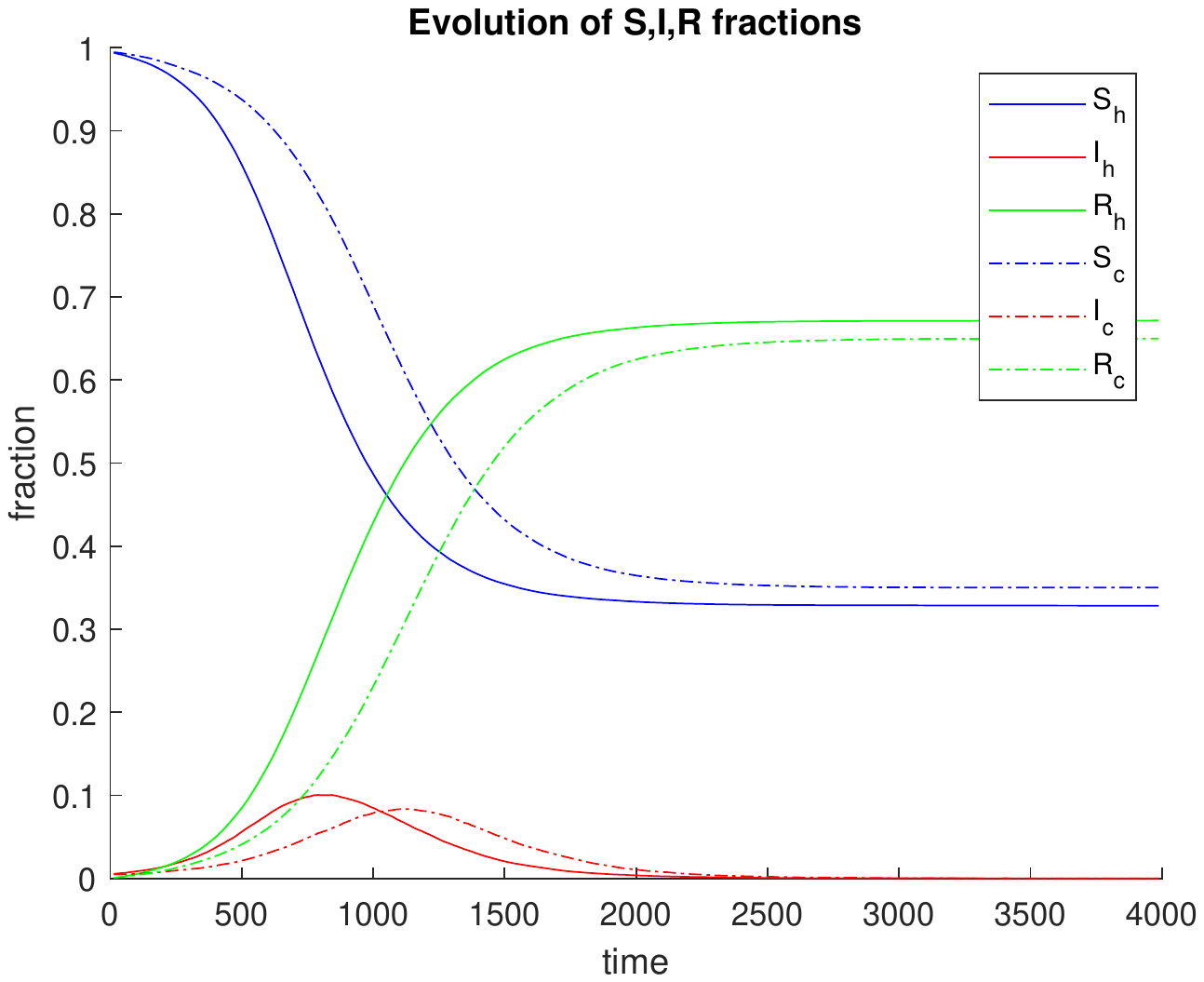}
}
\put(275,140){
  \includegraphics[width=0.38\textwidth]{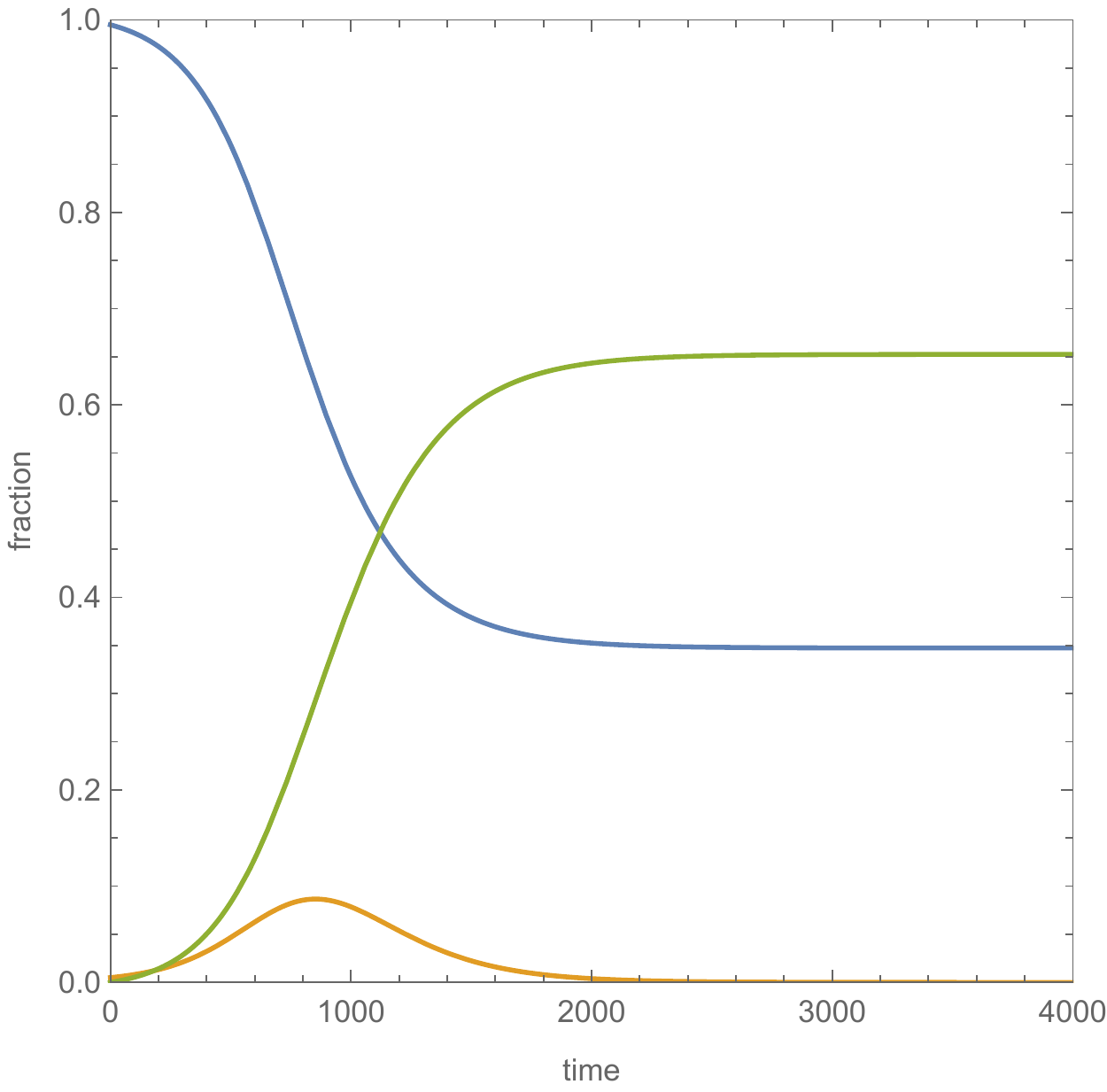}
}
\put(450,260){
  \includegraphics[width=0.06\textwidth]{legenda.png}
  }
\end{picture}
\caption{
Left panel: evolution of $S$, $I$, $R$  fractions for a particle system with hard sphere cross--section simulated via DSMC. The solid and dashed lines represent, respectively, the case of homogeneous and concentrated initial population of $I$ agents. Parameters: $N=180000$, $\b=0.75$, $\g=\frac{1}{120}$, $\lambda\simeq49.5$, $\tau\simeq55.85$.
Right panel: numerical solution of \eqref{eq:SIR'} with $m=\frac{1}{\tau}$, and the same $\b$ and $\g$ as in the left panel case.
}
\label{f:002}
\end{figure}

The case of hard sphere cross--section is reported in Fig.~\ref{f:002}. 
For uniform distributions (left panel, solid lines), we observe a small quantitative difference in the asymptotic fraction of the susceptible population (therefore of the recovered one) with respect to the solution of \eqref{eq:SIR'} (right panel).
Indeed, the DSMC tends to select colliding particles with large velocities: infected agents travelling with high speed are likely to transmit the infection.
This leads to a slightly wider diffusion of the $I$ population with respect to the system of ODEs \eqref{eq:SIR'} (in the experiment presented in Fig.~\ref{f:002}, $\bar{S}$ is estimated to be $0.318$ by the DSMC method while its actual value is $0.347$). 
 Moreover, we see that for a mean free path that is larger  compared to the size of the domain ($\lambda\simeq\frac{L}{20}$), starting from a concentrated initial datum does not change considerably the quantitative behaviour.

\begin{figure}[pht!]
\begin{picture}(200,170)(30,150)
\put(-20,-10){
  \includegraphics[width=0.75\textwidth]{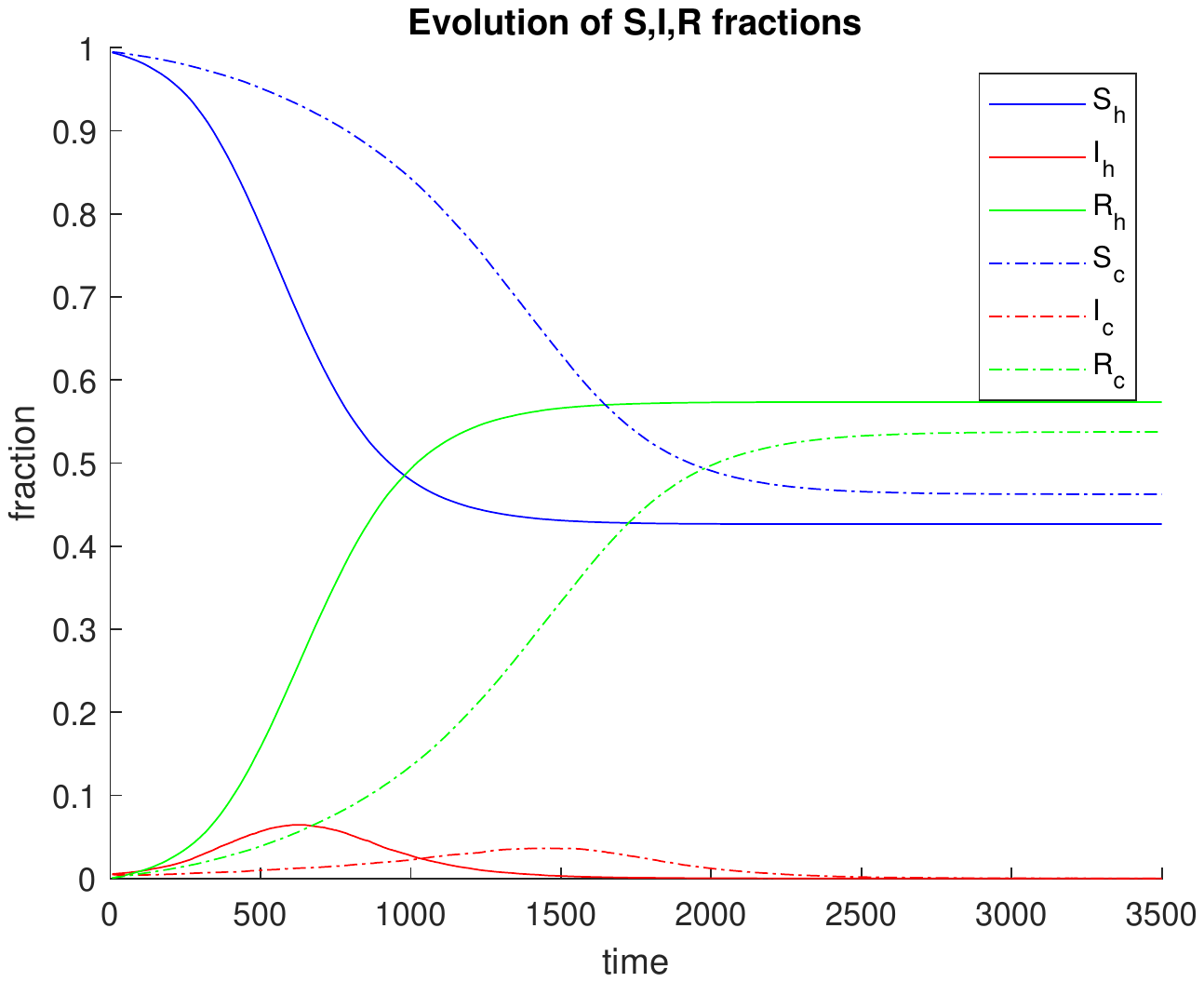}
}
\put(275,140){
  \includegraphics[width=0.38\textwidth]{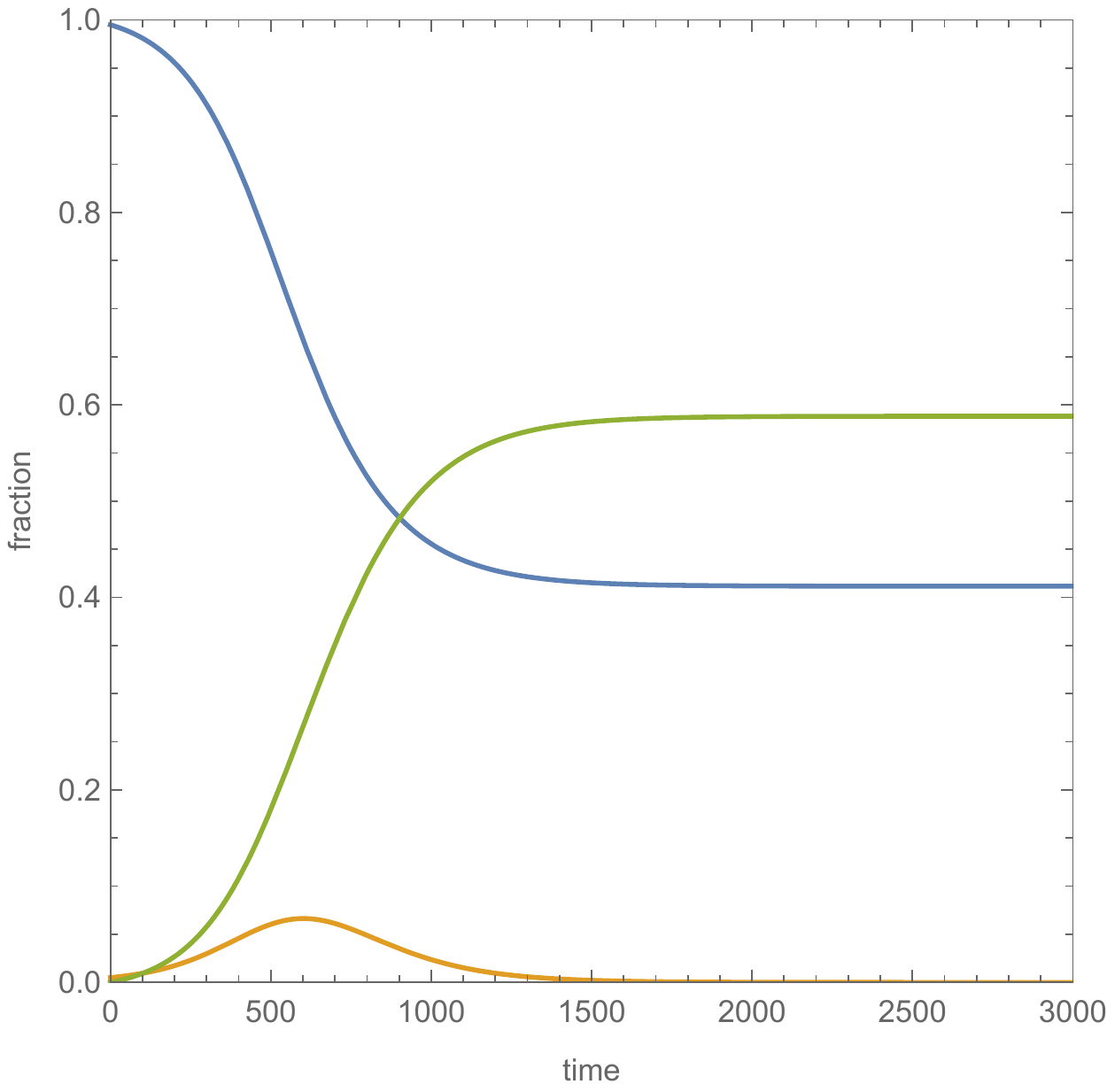}
}
\put(450,260){
  \includegraphics[width=0.06\textwidth]{legenda.png}
  }
\end{picture}
\caption{
Left panel: evolution of $S$, $I$, $R$  fractions for a particle system with semidiscrete cross--section simulated via DSMC. The solid and dashed lines represent, respectively, the case of homogeneous and concentrated initial population of $I$ agents. Parameters: $N=300000$, $\b=0.5$, $\g=\frac{1}{75}$, $\lambda\simeq25.0$, $\tau\simeq25.0$.
Right panel: numerical solution of \eqref{eq:SIR'} with $m=\frac{1}{\tau}$, and the same $\b$ and $\g$ as in the left panel case.
}
\label{f:003}
\end{figure}

Finally, the case of semidiscrete model cross-section is reported in Fig.~\ref{f:003}. 
We find excellent agreement between homogeneous particles system and SIR. 

\newpage
As expected, the larger asymptotic value of $S$ particles in the non-uniform cases is due to the time needed for the system to mix the populations,
the difference 
being more important for $\lambda$ small. 


\section{Meeting points} \label{sec:3}
\setcounter{equation}{0}    
\def\theequation{3.\arabic{equation}}

In this section we study three types of perturbation of the Boltzmann-SIR model, favouring non-equilibrium regimes. Symbolically, we call them ``supermarket'', ``airport'', and ``diffuse jet''.

\subsection{Supermarket} 
\label{ssec:super}

Let $D \subset \L $ be a box (the supermarket). In addition to the dynamics described by the free model, we assume that each particle jumps instantaneously in $D$ at an exponential time of rate $\g_1$. After the jump, the particle is uniformly distributed in $D$. Then it moves with unchanged velocity. The kinetic equations are:
\begin{equation}
\begin{cases}
  \left(\partial_t + v \cdot \nabla_x\right) f_S= Q(f_S,f)-\b Q_+(f_S,f_I) -\g_1 (f_S -\frac {\chi_D} {|D|}  g_S)\\
    \left(\partial_t + v \cdot \nabla_x\right) f_I = Q(f_I,f) +\b Q_+(f_S,f_I)-\g f_I -\g_1 (f_I -\frac {\chi_D} {|D|}  g_I)\\
     \left(\partial_t + v \cdot \nabla_x\right) f_R = Q(f_R,f)+\g f_I -\g_1 (f_R -\frac {\chi_D} {|D|}  g_R)\\
\end{cases}\;. \label{eq:SIR_BoltzmannS}
\end{equation}
Here $\chi_D$ is the characteristic function of $D$ and $|D|$ its area; $g_A$ with $A=S,I,R$ is the velocity distribution of $A$ i.e.\,$g_A(v)=\int dx \,f_A(x,v)$.
The equation for $f=\sum_A f_A$ is
\be
\label{trueBS}
\left(\partial_t + v \cdot \nabla_x\right) f= Q(f,f)-\g_1 (f -\frac {\chi_D} {|D|}  g)\nn
\ee
where $g$ is total velocity distribution $\int dx \, f$. The density $\r = \int dv\, f$ satisfies
$$
\left(\partial_t  \r + \text {div}_x(u\r)  \right)=-\g_1( \r- \frac {\chi_D} {|D|})
$$
where $\r u = \int dv\,v\, f$.
As we are not able to characterize explicitly the stationary solutions to Eq.\,\eqref{eq:SIR_BoltzmannS}, we turn to numerical investigation.

\begin{figure}[ht!]
\begin{picture}(200,170)(30,150)
\put(-20,-10){
  \includegraphics[width=0.70\textwidth]{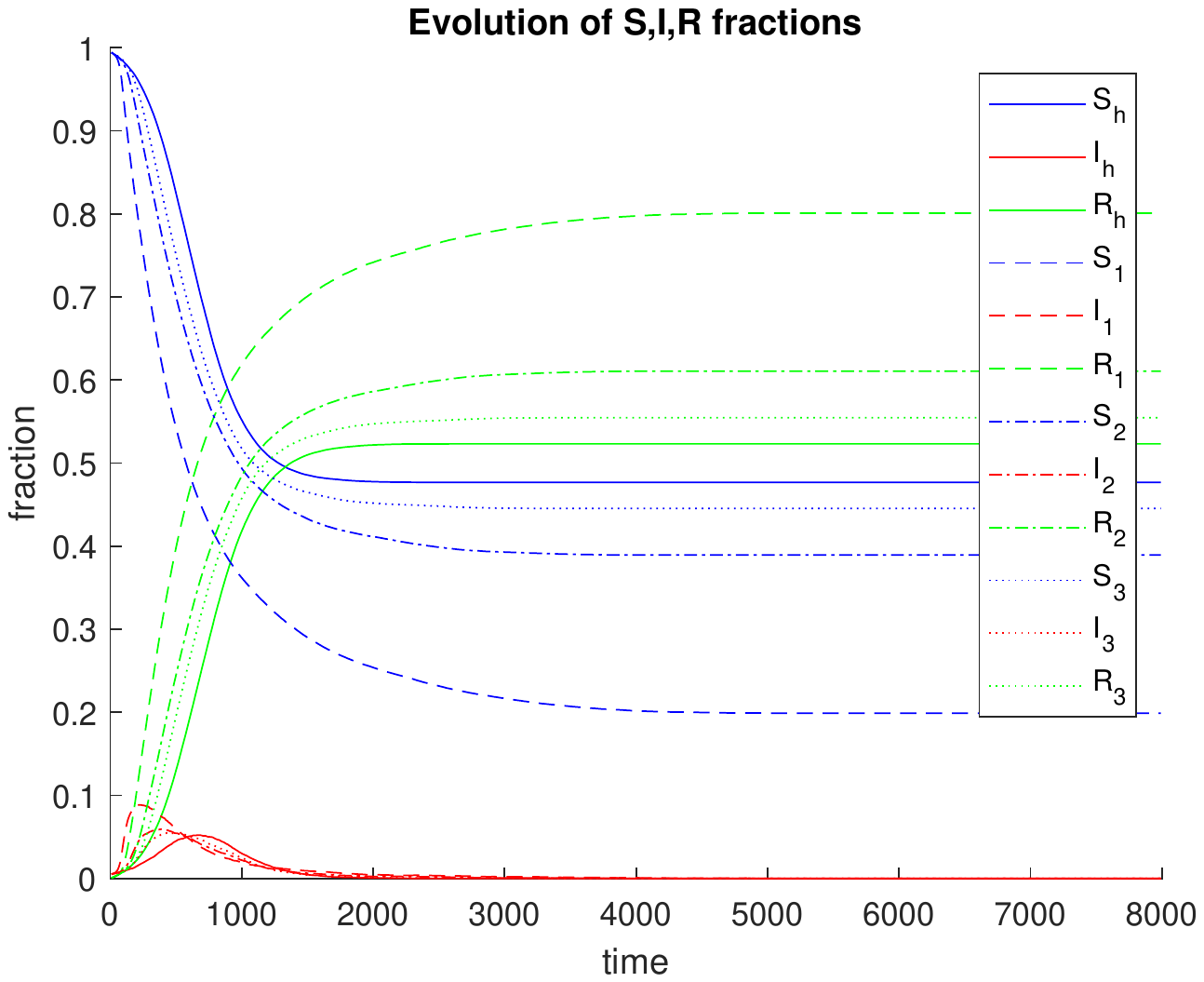}
}
\put(205,-10){
  \includegraphics[width=0.70\textwidth]{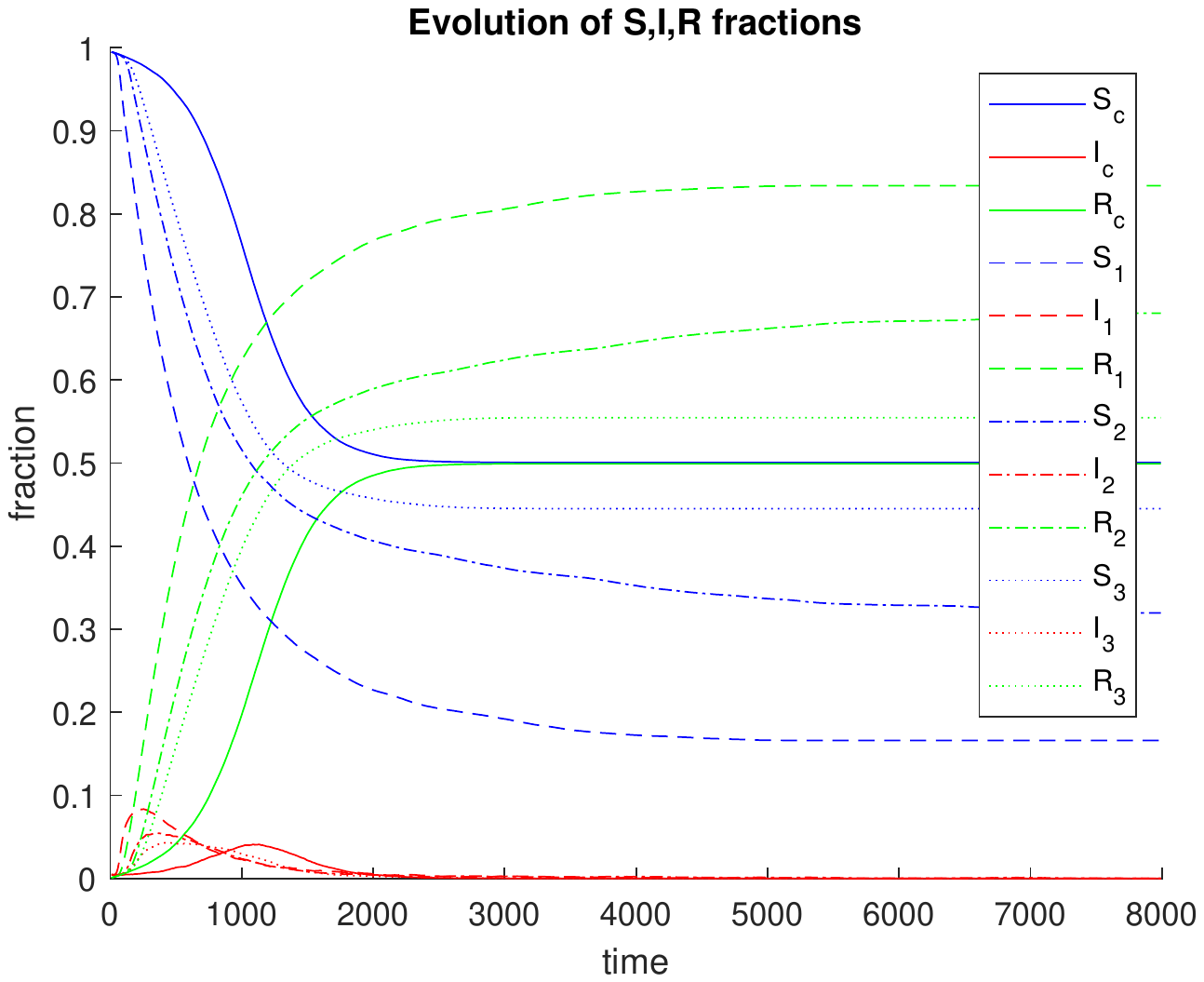}
}
\end{picture}
\caption{
Evolution of $S$, $I$, $R$  fractions for a particle system with hard sphere cross--section simulated via DSMC, Eq.\,\eqref{eq:SIR_BoltzmannS}. 
Parameters: $N=180000$, $\b=0.7$, $\g=\frac{1}{75}$, $\lambda\simeq33.5$, $\tau\simeq37.8$.
Left panel: $|D|=\frac{1}{100}|\Lambda|$, homogeneous initial datum. Subscript $h$ refers to the free model (no jumps), subscript $i$ refers to assumed value $\g_1=(i\cdot 10^3)^{-1}$, $i=1,2,3$.
Right panel: $|D|=\frac{6}{1000}|\Lambda|$, concentrated initial datum. 
}
\label{f:004}
\end{figure}

The local higher density (in $D$, and in a neighbourhood of it) 
makes more likely the increase of $I$ particles.
Ultimately, this leads to an asymptotic behaviour with lower number of susceptible agents (with respect to the free model).
Two parameters contribute to magnify the effect: the intensity of jumps $\g_1$ and the smallness of the box $|D|$, as reported in Fig~\ref{f:004}. 
Note that here a smaller $D$ produces a more significant effect on the difference in the asymptotics with respect to the different initial configurations.
This is at variance with the free model,  where uniform initial data provide a wider diffusion than a concentrated initial population $I$.

%
%
%

\subsection{Airport}

In addition to a density concentration, we consider now the action of an external flow.
As before, agents $S$ and $R$ jump in $D$ with a rate $\g_1$ (infects are not allowed to fly). Then they disappear and are simultaneously replaced by an equal number of agents. The injected agents are either $S$ or $R$, with equal probability $(1-\a)/2$, or $I$ with probability $\a$. The extreme case is $\a=1$ (maximal flux of infects). The equations are\footnote{
An immediate generalization of Eq.\,\eqref{eq:SIR_BoltzmannA} is obtained by considering different fractions $\a_1$, $\a_2$ and $\a_3$ (with $\a_1+\a_2+\a_3=1$) in place of $\a$, $\frac{1-\a}{2}$ and $\frac{1-\a}{2}$. 
This does not change the qualitative behaviour.
%
}:
\begin{equation}
\begin{cases}
  \left(\partial_t + v \cdot \nabla_x\right) f_S= Q(f_S,f)-\b Q_+(f_S,f_I) -\g_1  (f_S - (1-\a) \frac {\chi_D} {2|D|}  (g_S+g_R))\\
    \left(\partial_t + v \cdot \nabla_x\right) f_I = Q(f_I,f) +\b Q_+(f_S,f_I) -\g f_I + \g_1 \a \frac {\chi_D} {|D|}  (g_S+g_R)\\
     \left(\partial_t + v \cdot \nabla_x\right) f_R = Q(f_R,f)+\g f_I -\g_1 (f_R -(1-\a) \frac {\chi_D} {2|D|}  (g_S+g_R))\\
\end{cases}\;. \label{eq:SIR_BoltzmannA}
\end{equation}
In this case, $f$ does not solve a closed equation.

\begin{figure}[ht!]
\begin{picture}(200,170)(30,150)
\put(-20,-10){
  \includegraphics[width=0.70\textwidth]{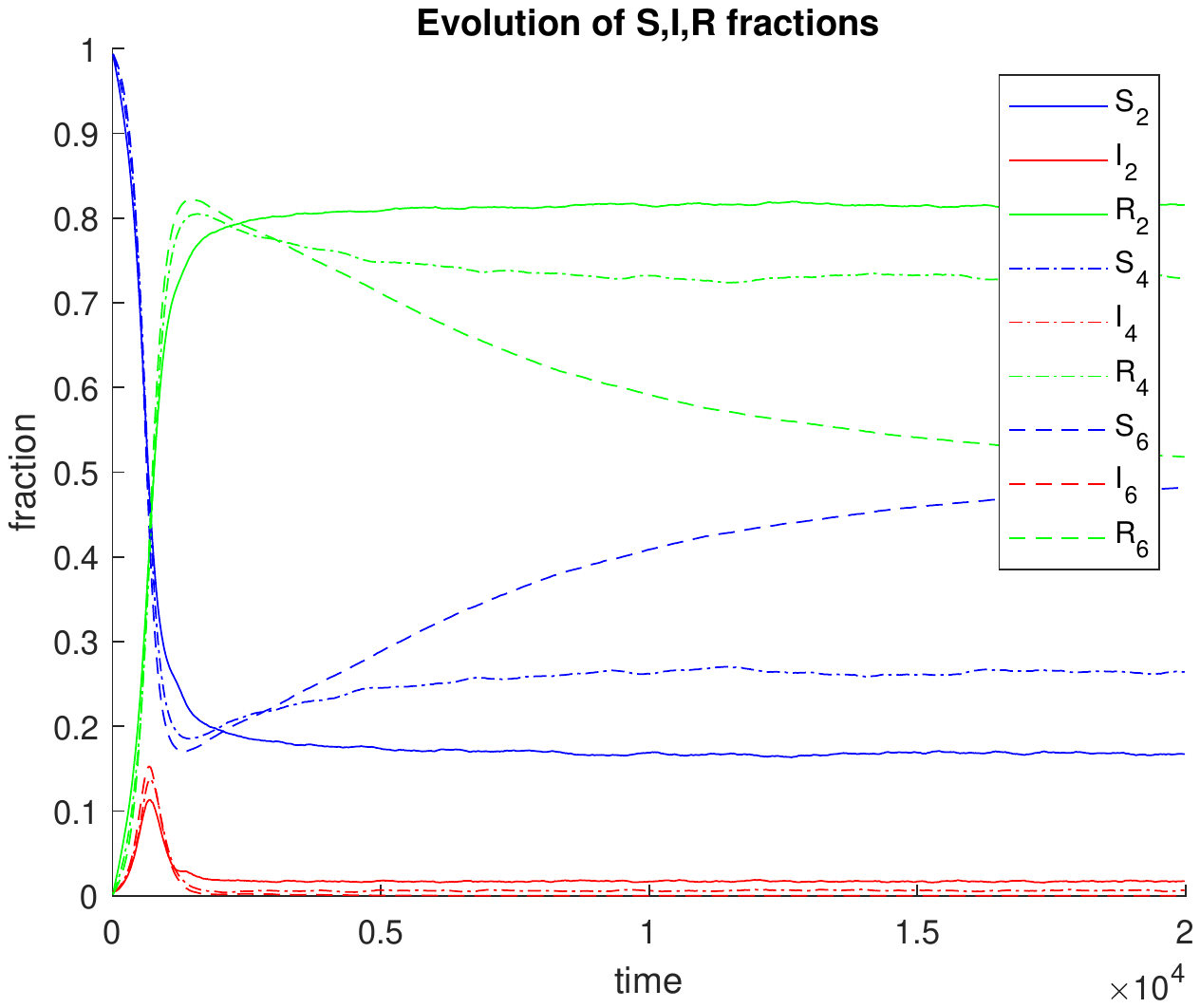}
}
\put(205,-10){
  \includegraphics[width=0.70\textwidth]{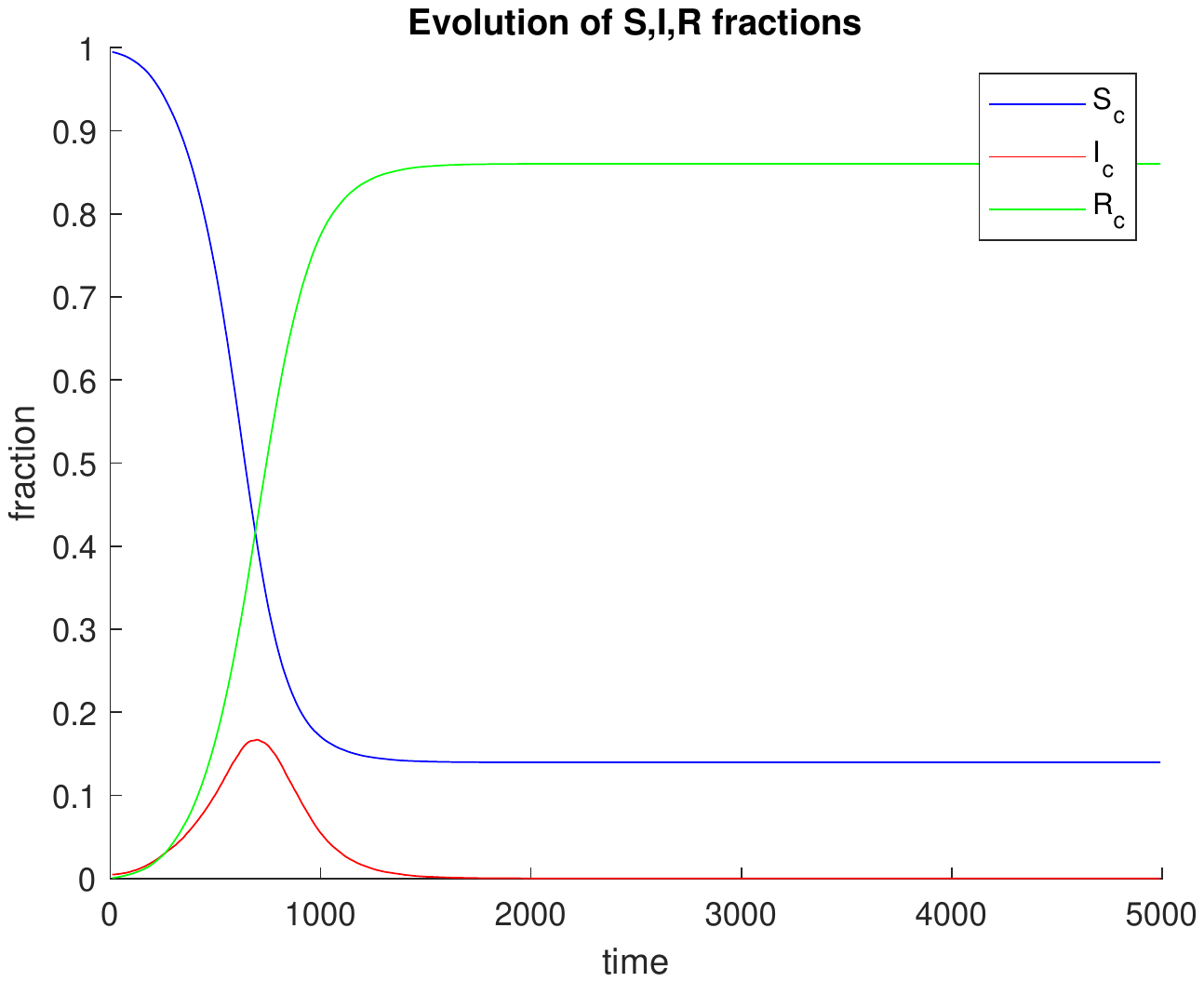}
}
\end{picture}
\caption{Evolution of $S$, $I$, $R$  fractions for a particle system with hard sphere cross-section simulated via DSMC, Eq.\,\eqref{eq:SIR_BoltzmannA}. 
Parameters: $N=180000$, $\b=0.9$, $\g=\frac{1}{100}$, $\lambda\simeq33.4$, $\tau\simeq37.7$, $|D|=\frac{1}{100}|\Lambda|$, $\a=0$, concentrated initial datum. 
 Left panel: subscript
 $i$ refers to assumed value $\g_1=(i\cdot 10^3)^{-1}$, $i=2,4,6$.
 Right panel: free model ($\g_1=0$), same parameters.
 }
\label{f:005}
\end{figure}

\begin{figure}[ht!]
\begin{picture}(200,170)(30,150)
\put(-20,-10){
  \includegraphics[width=0.70\textwidth]{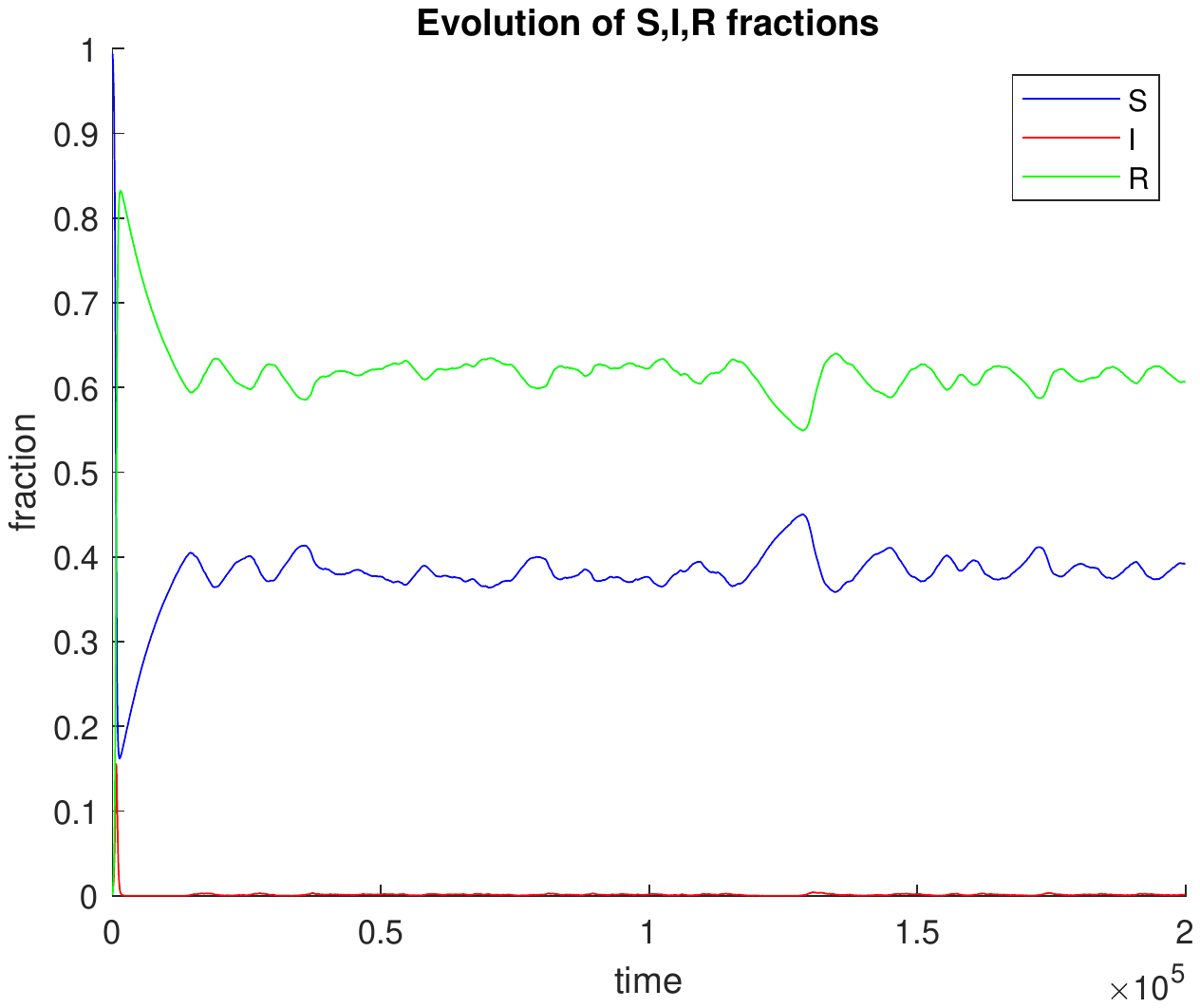}
}
\put(205,-10){
  \includegraphics[width=0.70\textwidth]{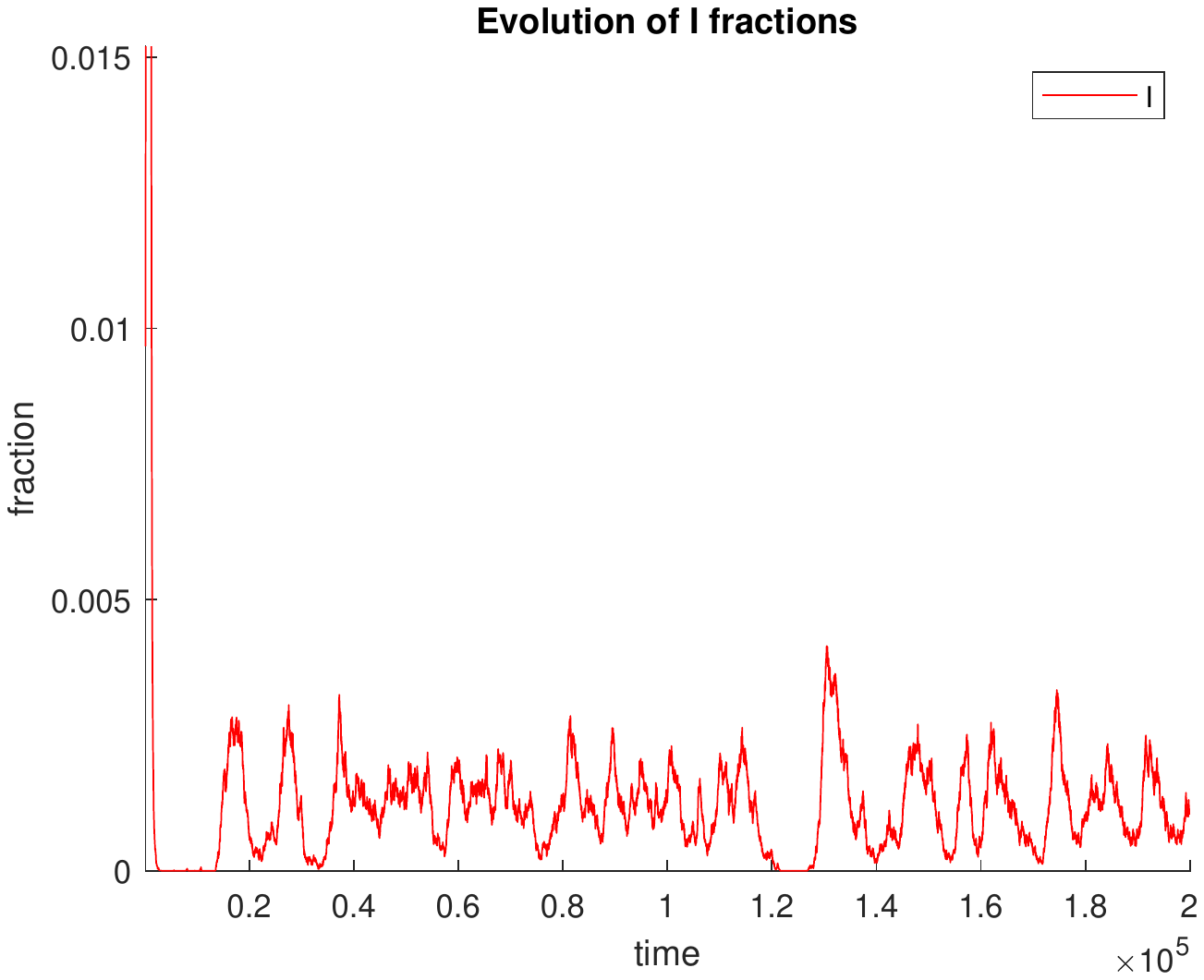}
}
\end{picture}
\caption{
Particle system with hard sphere cross-section simulated via DSMC, Eq.\,\eqref{eq:SIR_BoltzmannA}. 
Parameters: $N=180000$, $\b=0.9$, $\g=\frac{1}{100}$, $\lambda\simeq33.4$, $\tau\simeq37.7$, $|D|=\frac{1}{100}|\Lambda|$, $\a=2\cdot 10^{-5}$,  $\g_1=10^{-4}$, concentrated initial datum. 
 Left panel: Evolution of $S$, $I$, $R$  fractions.
 Right panel: detail of the $I$ fraction.
 }
\label{f:006}
\end{figure}

\begin{figure}[ht!]
\begin{picture}(200,170)(30,150)
\put(-20,-10){
  \includegraphics[width=0.70\textwidth]{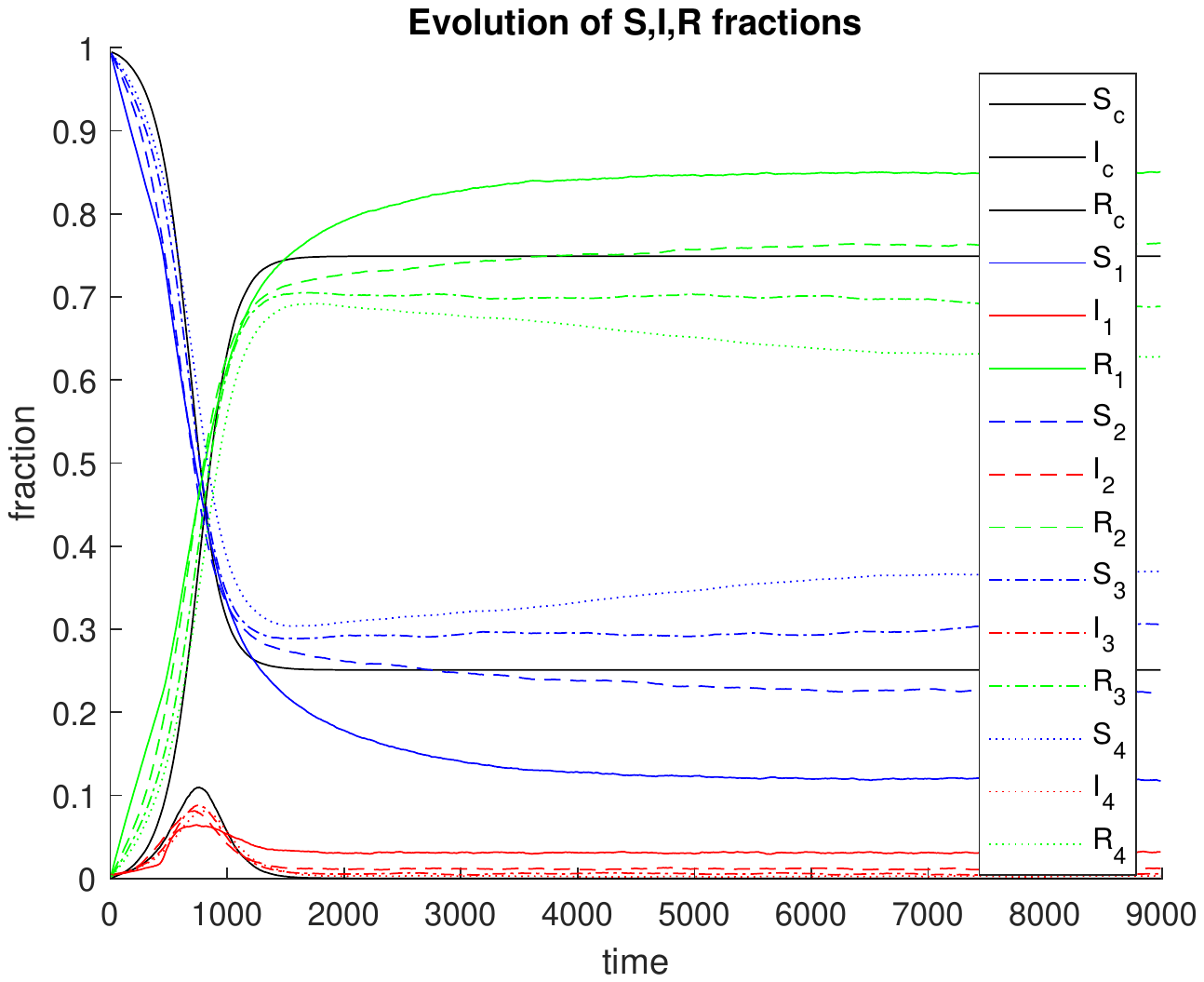}
}
\put(205,-10){
  \includegraphics[width=0.70\textwidth]{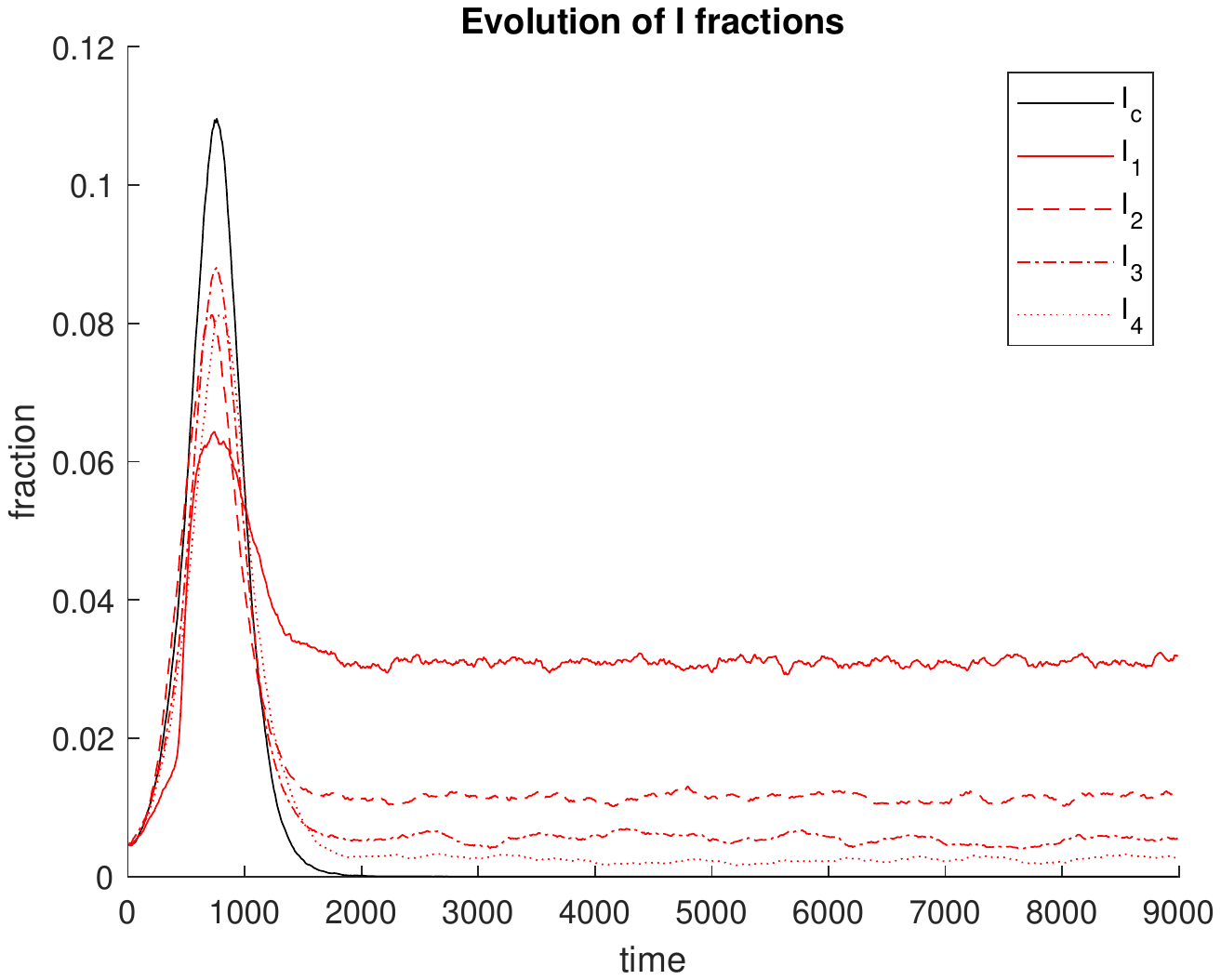}
}
\end{picture}
\caption{
Particle system with hard sphere cross--section simulated via DSMC, Eq.\,\eqref{eq:SIR_BoltzmannA}. 
Parameters: $N=180000$, $\b=0.9$, $\g=\frac{1}{80}$, $\lambda\simeq33.4$, $\tau\simeq37.7$, $|D|=\frac{1}{100}|\Lambda|$, $\a=10^{-4}$, concentrated initial datum. 
Subscript 
$c$ refers to the free model ($\g_1=0$), subscript
 $i$ refers to assumed value $\g_1=(i\cdot 10^3)^{-1}$, $i=1,2,3,4$.
 Left panel: Evolution of $S$, $I$, $R$  fractions.
 Right panel: detail of the $I$ fraction.
 }
\label{f:007}
\end{figure}

The evolution shows two time scales. 
In a first  phase, it is qualitatively close to the free model. This is true until a herd immunity threshold,
after which the injection of $I$ particles in $D$ crucially determines the long time behaviour.
In the extreme case $\a=1$, no susceptible agent survives. 
In the opposite case $\alpha=0$ (no infected are ever introduced), 
if the jump rate is sufficiently intense, the infection may  be never extinguished: the source of susceptible agents
leads to a stationary configuration where all three populations are non--zero; see  Fig.~\ref{f:005}, $\g_1=(i\cdot 10^3)^{-1}$, $i=2,4$.
If  instead the jump rate is low, the asymptotic values are $\bar{S}=\bar{R}=0.5$ when all the agents have jumped at least once after the extinction of $I$ (Fig.~\ref{f:005}, $\g_1=(6\cdot 10^3)^{-1}$). 
%
If, additionally, a small fraction of agents $I$ is injected,
recurrent small waves arise; see Figures~\ref{f:006} and \ref{f:007}, corresponding respectively to cases with extinction and without extinction of $I$ (depending on $\a, \g_1$).


%

%

%
%

\newpage
\subsection{Diffuse jets}

Finally, we consider the effect of an external flow, as in the previous section, but without density localization. This corresponds to simple random diffuse replacement of non infected agents by infected agents with probability $\a$, and by susceptible and recovered agents with equal probability $\frac {1-\a}2$: 
\begin{equation}
\begin{cases}
  \left(\partial_t + v \cdot \nabla_x\right) f_S= Q(f_S,f)-\b Q_+(f_S,f_I) -\g_1  (f_S - \frac {(1-\a)}2  (f_S+f_R))\\
    \left(\partial_t + v \cdot \nabla_x\right) f_I = Q(f_I,f) +\b Q_+(f_S,f_I) -\g f_I + \g_1 \a (f_S+f_R)\\
     \left(\partial_t + v \cdot \nabla_x\right) f_R = Q(f_R,f)+\g f_I -\g_1 (f_R - \frac {(1-\a)}2  (f_S+f_R))\\
\end{cases}\;. \label{eq:SIR_BoltzmannDA}
\end{equation}
%
%
%
In this case, one has a close system of equations for averaged fractions (holding in case of homogeneous solutions for Maxwellian molecules, or after thermalization), reminiscent of SIR-like models with more possible reactions:
\be \label{eq:SIR:jet}
\begin{cases}
  \dot S = -\b m IS + \g_1 \left(- \frac {1+\a}2 S + \frac {1-\a}2 R\right)  \\
    \dot I = \b m IS -\g I+\g_1 \a (S+R) \\
     \dot R = \g I + \g_1 \left(- \frac {1+\a}2 R + \frac {1-\a}2  S\right)\\
\end{cases}\;.
\ee
For $m=\g_1=1$, it reduces to 
\be \label{eq:SIR:jet'}
\begin{cases}
  \dot S = -\b  IS+ \frac {R-S}2  - \frac {\a}2 (S+R)  \\
    \dot I = \b  IS -\g I+\a (S+R) \\
     \dot R = \g I + \frac {S-R}2  - \frac {\a}2 (S+R)\\
\end{cases}\;, \nn
\ee
where it is easier to recognize the competing terms effect.

For $\a =1$ we have, again, extinction of the susceptible agents; for $\a \neq1$ non--trivial stationary solutions exist for the three populations.
In the case $\a=0$, for $\g_1$ sufficiently small, the simulation of \eqref{eq:SIR_BoltzmannDA} 
displays an instant of total vanishing of the infection: 
$I(t)$ is equal to $0$ from this time on, and the asymptotic values are $\bar{S}=\bar{R}=\frac{1}{2}$. 
The ODE system \eqref{eq:SIR:jet}, instead, has solution $I(t)\neq 0$ for every time $t\geq0$. The asymptotic values are here close to the case $\a \neq 0$ very small and the values $\bar{S}$ and $\bar{R}$ are not $1/2$.

\begin{figure}[pth!]
\begin{picture}(200,370)(30,150)
  \put(-20,180){
  \includegraphics[width=0.75\textwidth]{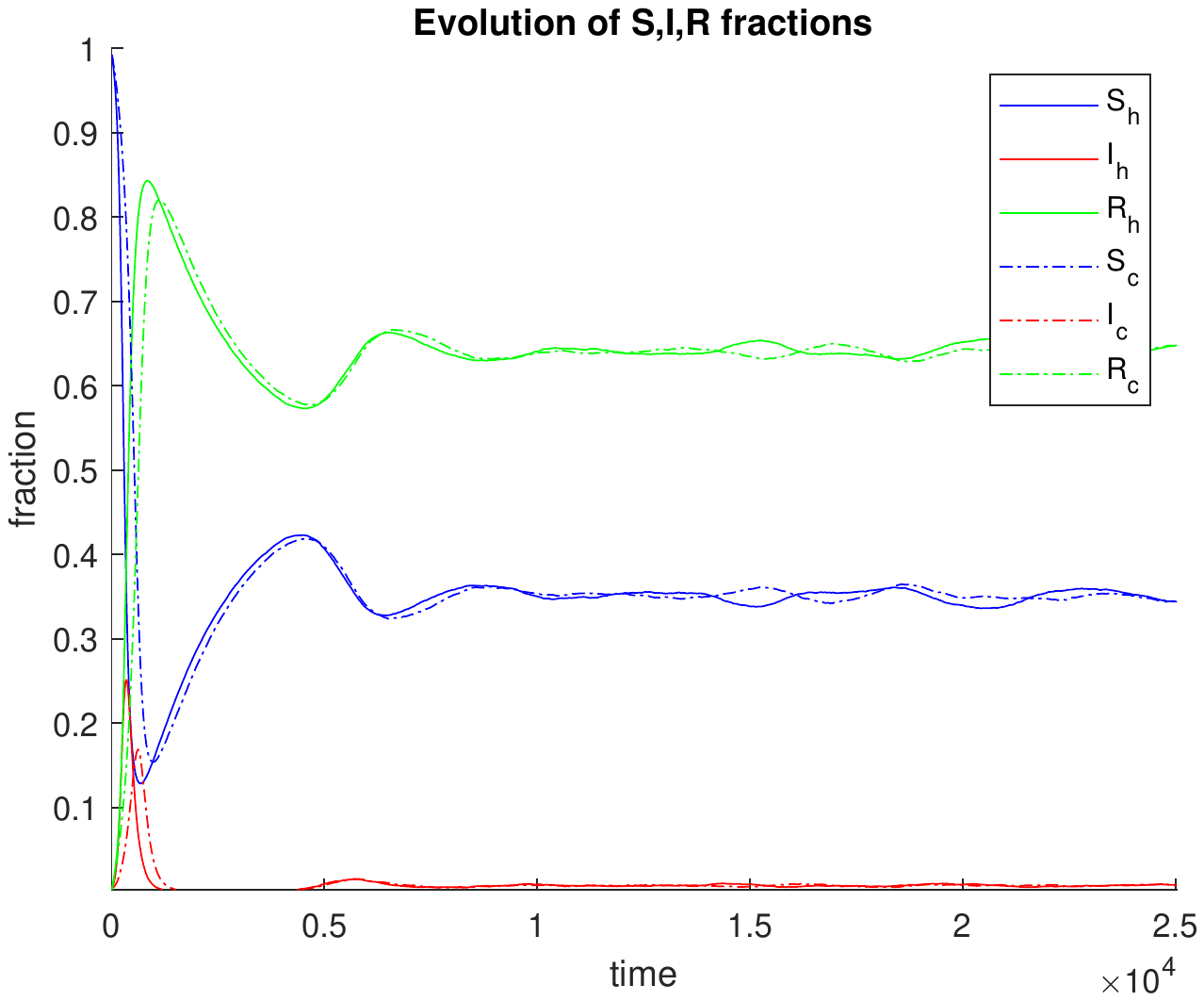}
}
\put(275,333){
  \includegraphics[width=0.38\textwidth]{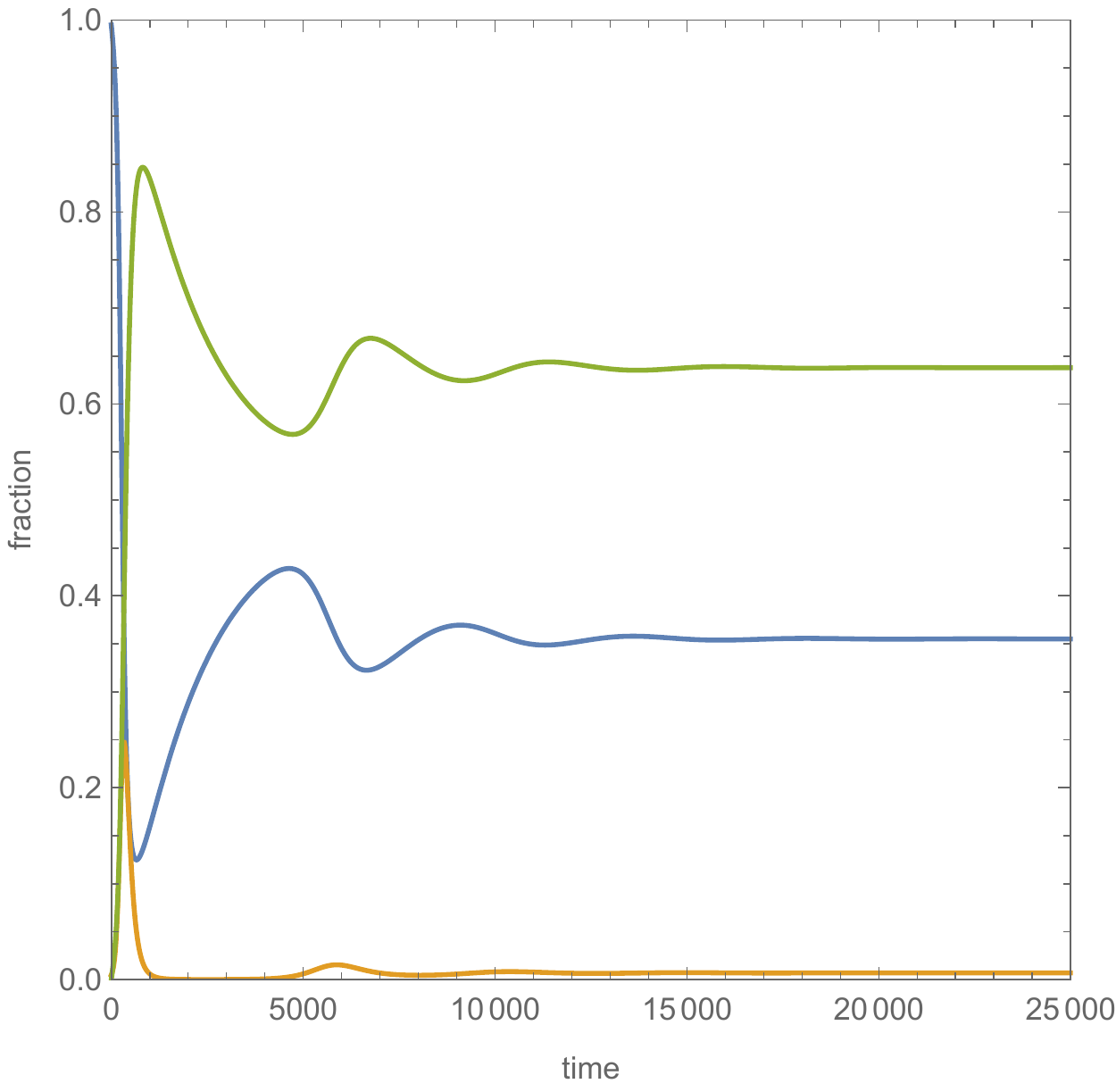}
}
\put(450,450){
  \includegraphics[width=0.06\textwidth]{legenda.png}
  }
\put(-20,-10){
  \includegraphics[width=0.75\textwidth]{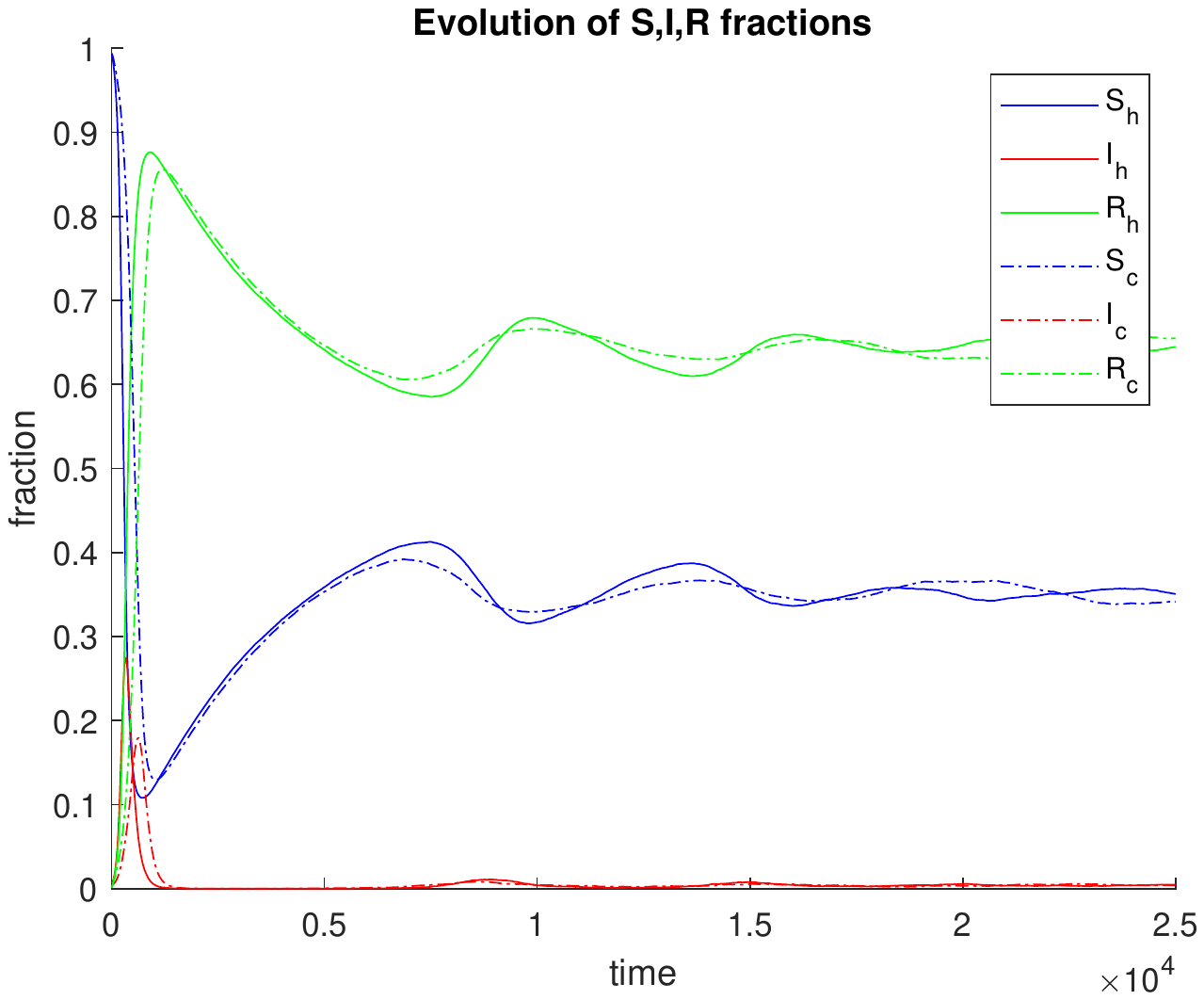}
}
\put(275,143){
  \includegraphics[width=0.38\textwidth]{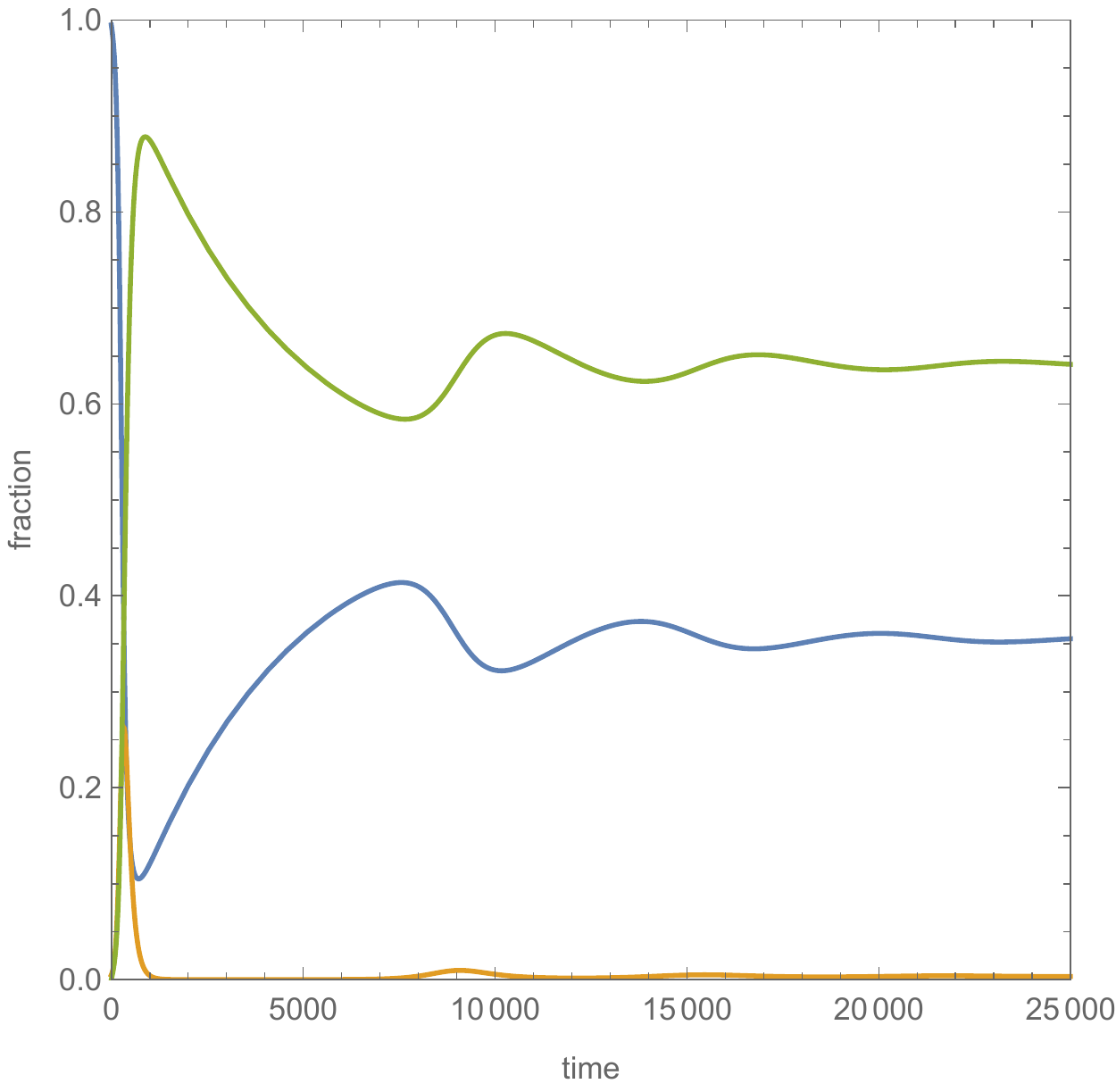}
}
\put(450,260){
  \includegraphics[width=0.06\textwidth]{legenda.png}
  }
\end{picture}
\caption{Evolution of $S$, $I$, $R$  fractions. Top panels: $\g_1=\frac 1 {2000}$, bottom panels $\g_1=\frac 1 {4000}$.
Left panel:  particle system with hard spheres cross--section simulated via DSMC. The solid and dashed lines represent, respectively, the case of homogeneous and concentrated initial population of $I$ agents. Parameters: $N=200000$, $\b=0.95$, $\g=\frac{1}{100}$, $\lambda\simeq29.9$, $\tau\simeq33.8$,  $\a=2\cdot 10^{-4}$.
Right panel: numerical solution of \eqref{eq:SIR:jet} for $m=\frac{1}{\tau}$, and the same $\b$, $\g$,  $\g_1$, $\a$ as the particle system.
}
\label{f:008}
\end{figure}

In line with previous results in Section \ref{ssec:simul_free}, 
%
the qualitative behaviour is well captured by the ODE system \eqref{eq:SIR:jet}, as reported in Figure \ref{f:008}.
After a mixing time, the system is totally homogeneous (the perturbation itself being homogeneous).
We show in Fig.~\ref{f:008} some experiments exhibiting damped waves, comparing both homogeneous and concentrated initial data for the particle system (left panels) with the solution of \eqref{eq:SIR:jet} (right panels).
%

\begin{figure}[pth!]
\begin{picture}(200,542)(-10,10)
  \put(150,0){
  \includegraphics[width=0.27\textwidth]{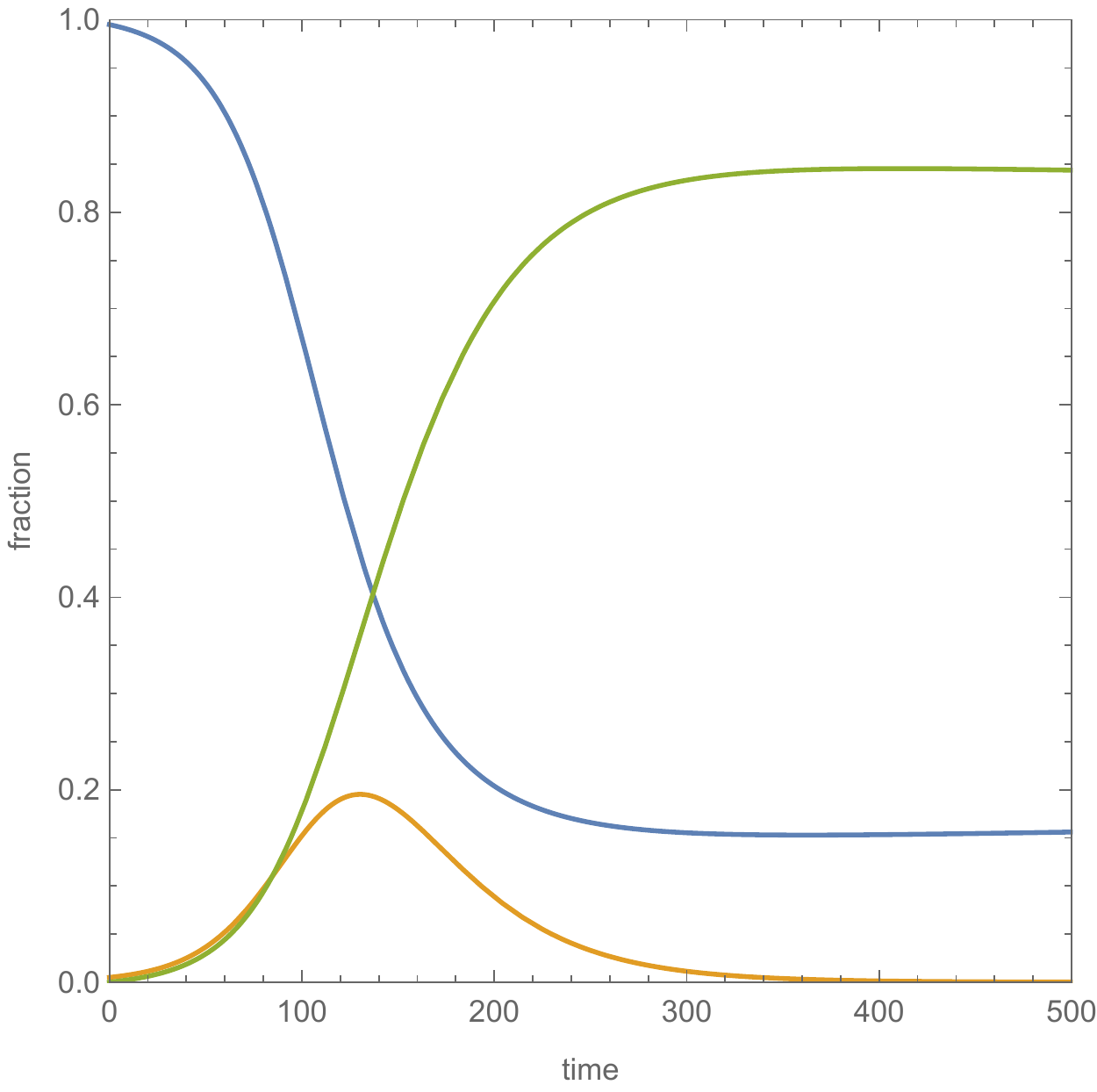}
}
\put(320,0){
  \includegraphics[width=0.27\textwidth]{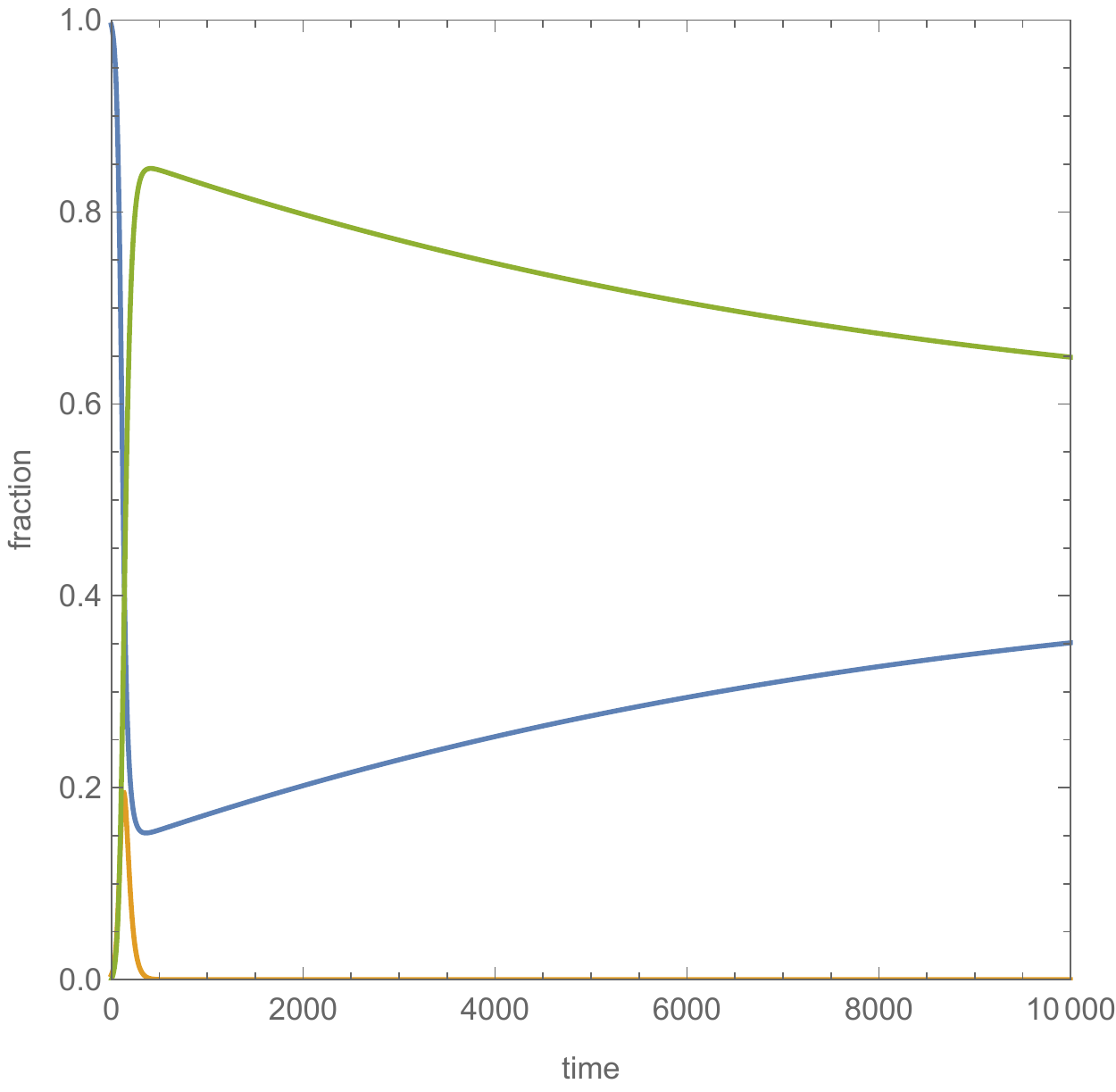}
}
\put(0,0){
  \includegraphics[width=0.25\textwidth]{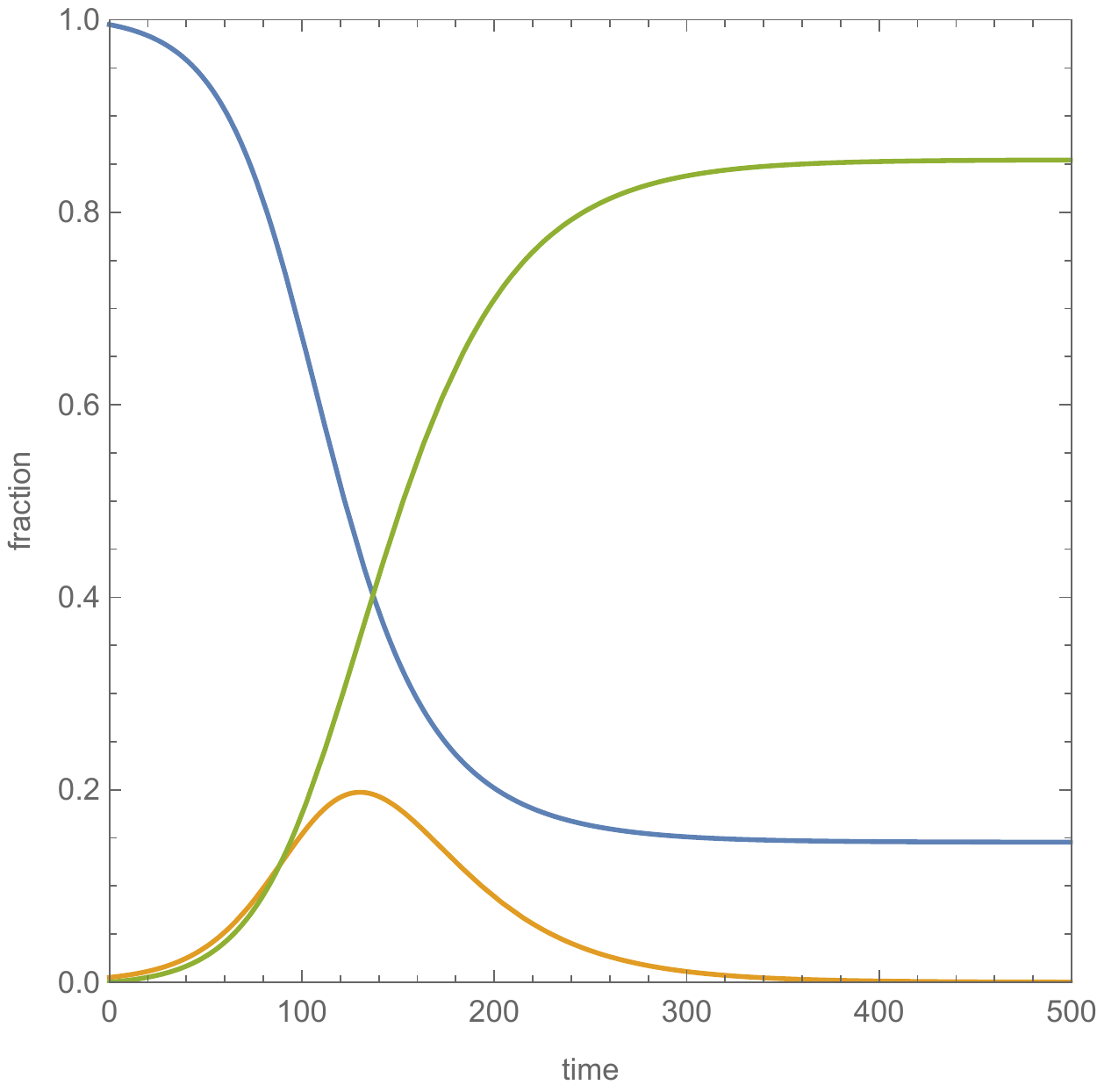}
}
\put(0,125){
  \includegraphics[width=0.25\textwidth]{last_sir.pdf}
}\put(0,250){
  \includegraphics[width=0.25\textwidth]{last_sir.pdf}
}\put(0,375){
  \includegraphics[width=0.25\textwidth]{last_sir.pdf}
}
\put(0,500){
  \includegraphics[width=0.25\textwidth]{last_sir.pdf}
}
  \put(150,125){
  \includegraphics[width=0.27\textwidth]{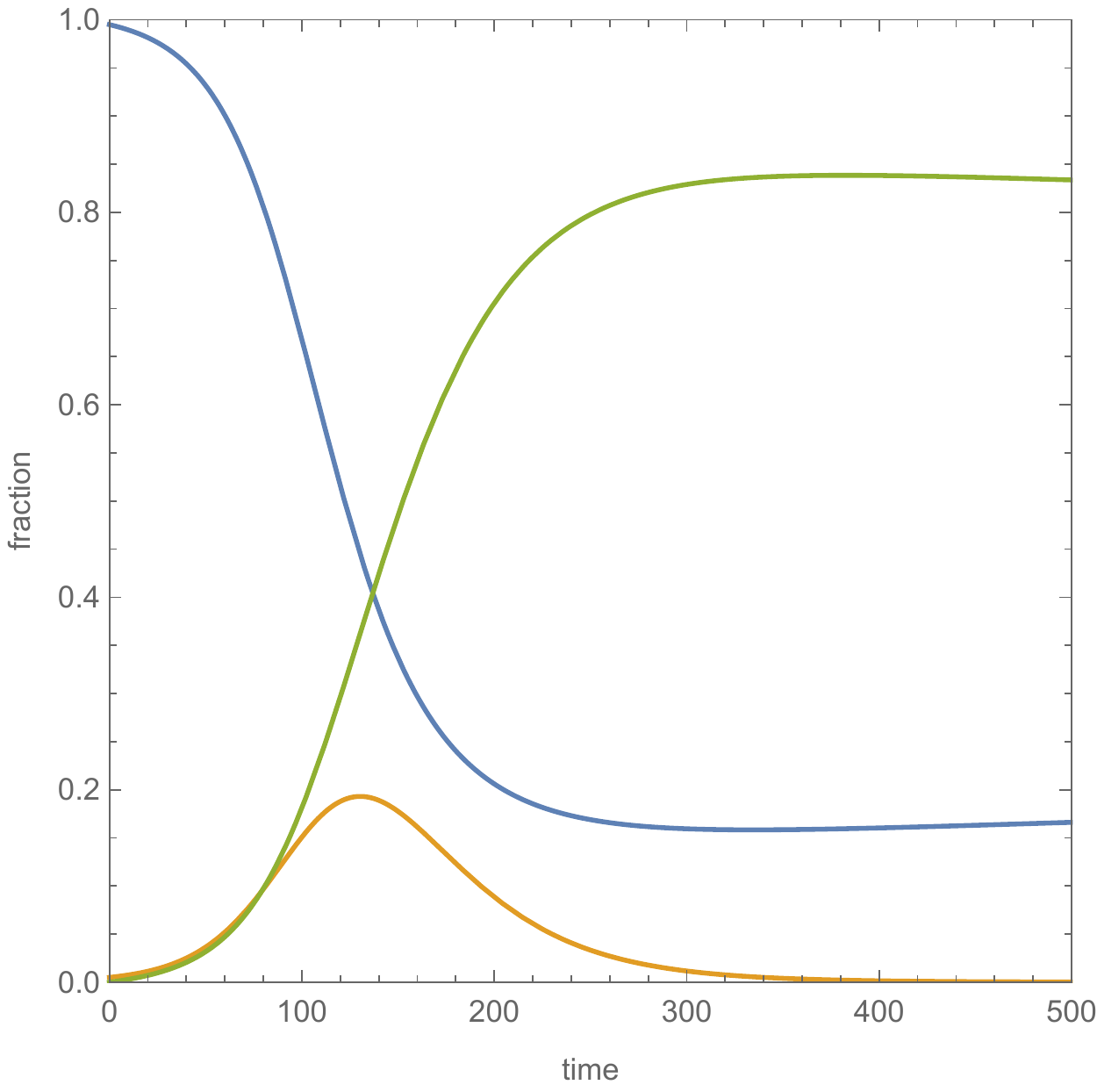}
}
\put(320,125){
  \includegraphics[width=0.27\textwidth]{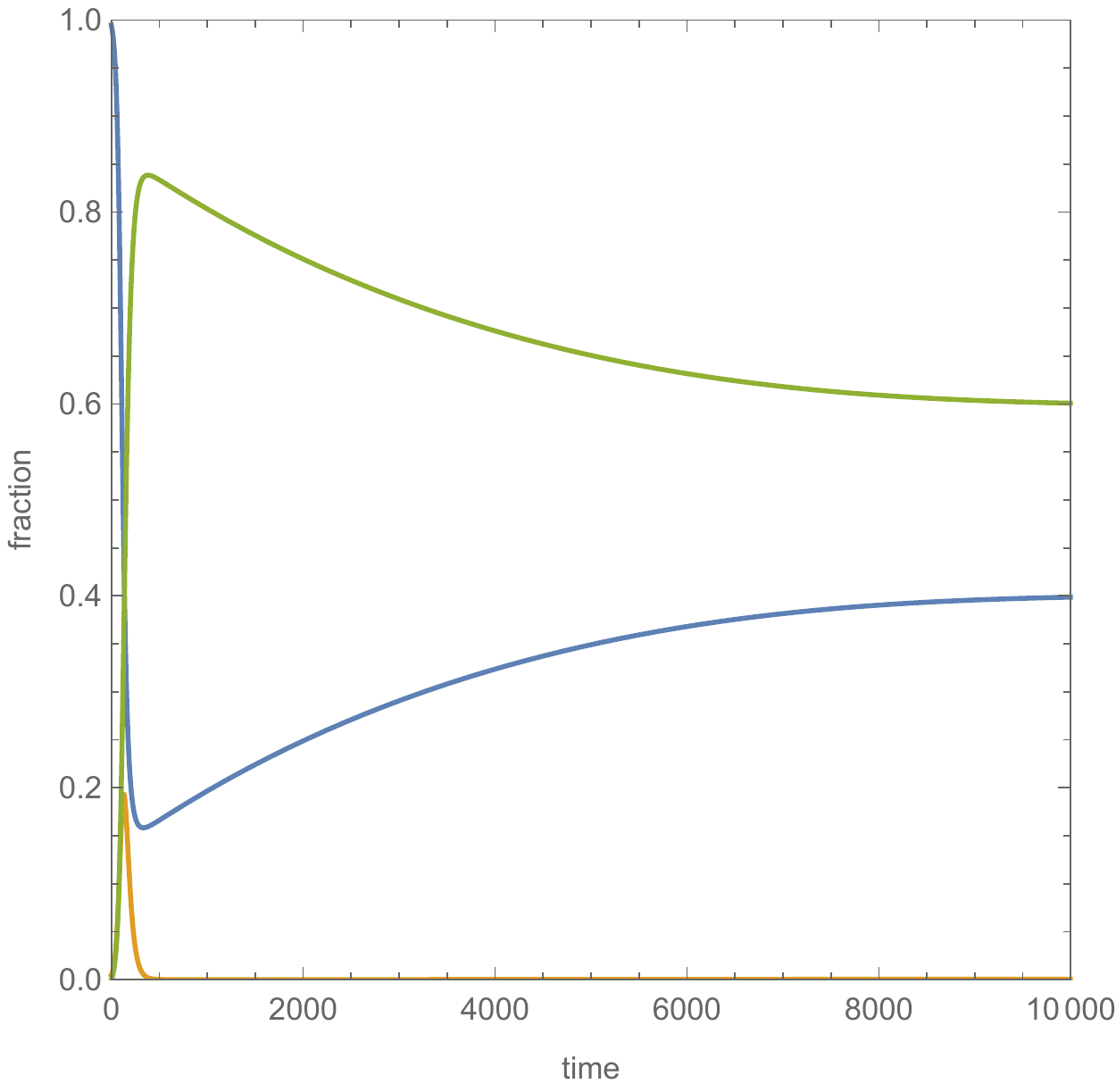}
}
  \put(150,250){
  \includegraphics[width=0.27\textwidth]{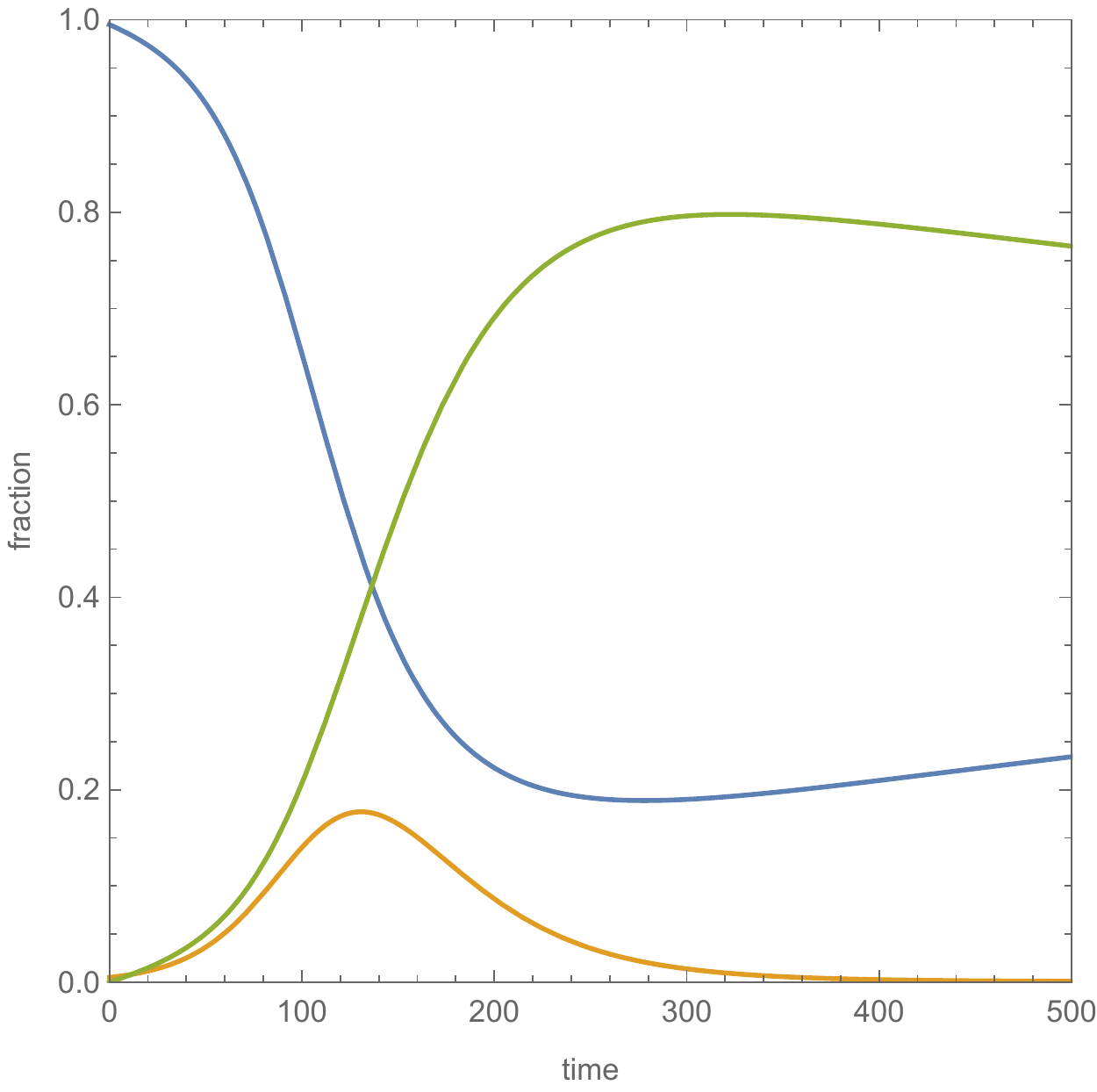}
}
\put(320,250){
  \includegraphics[width=0.27\textwidth]{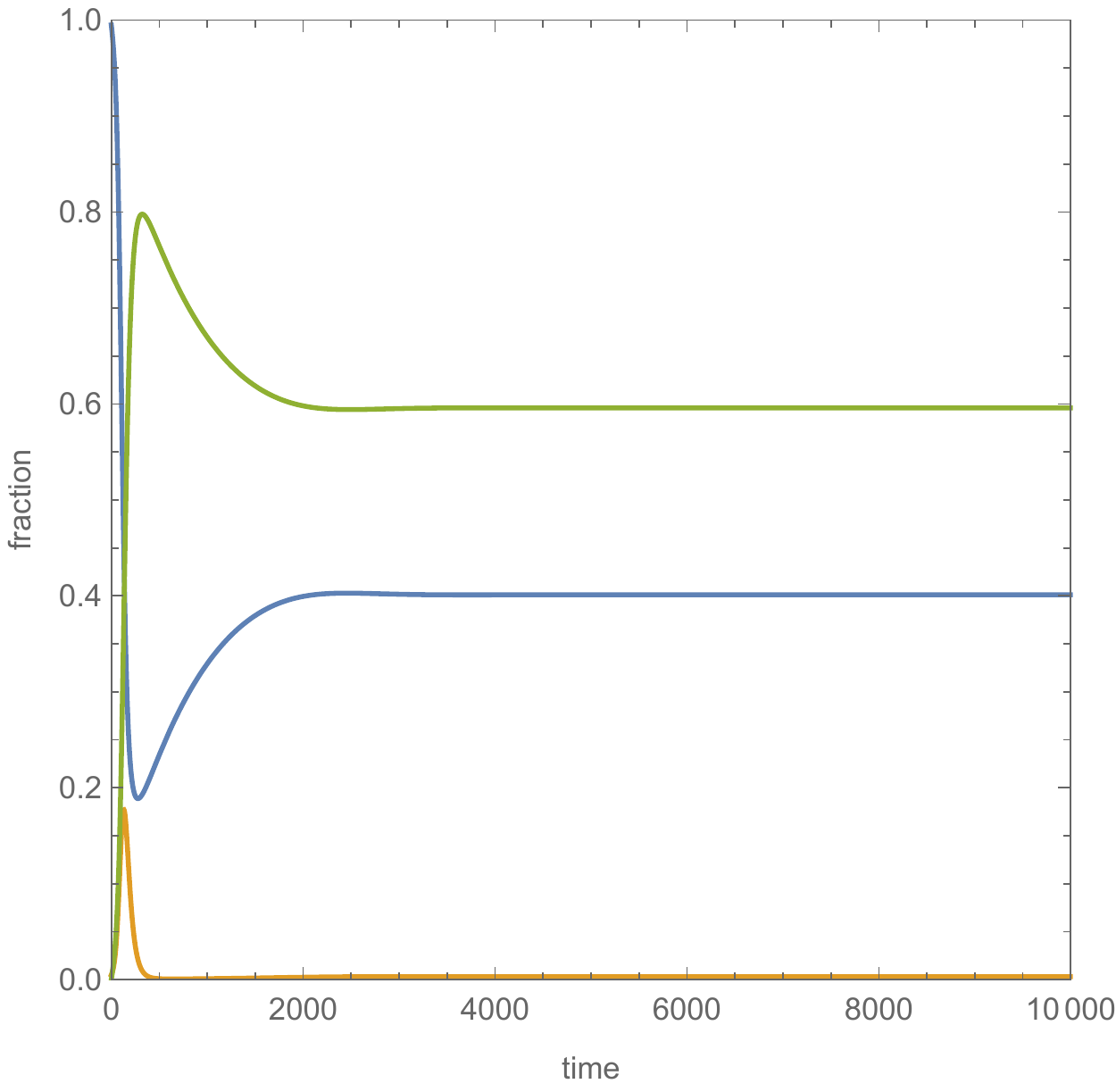}
}
  \put(150,375){
  \includegraphics[width=0.27\textwidth]{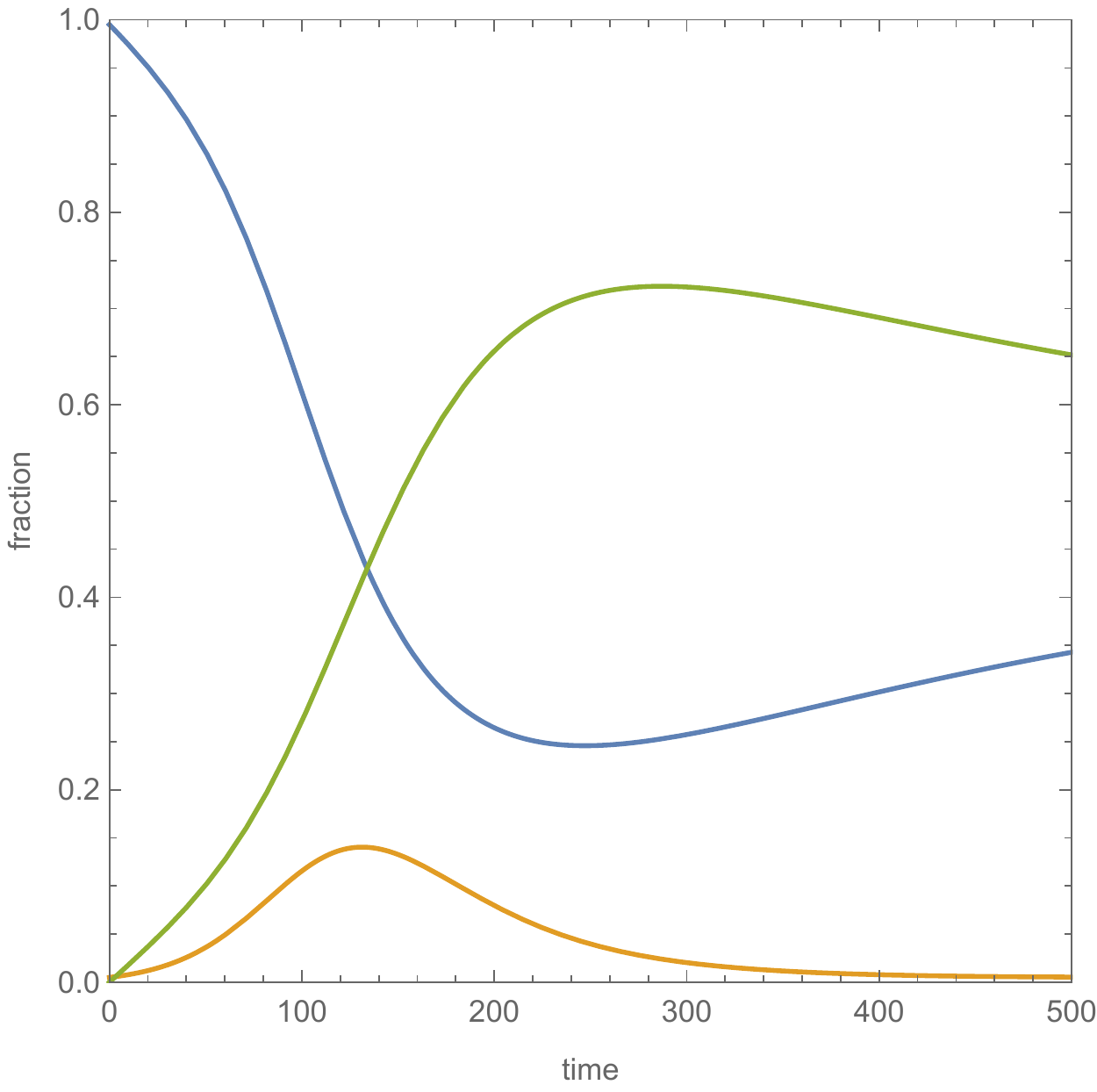}
}
\put(320,375){
  \includegraphics[width=0.27\textwidth]{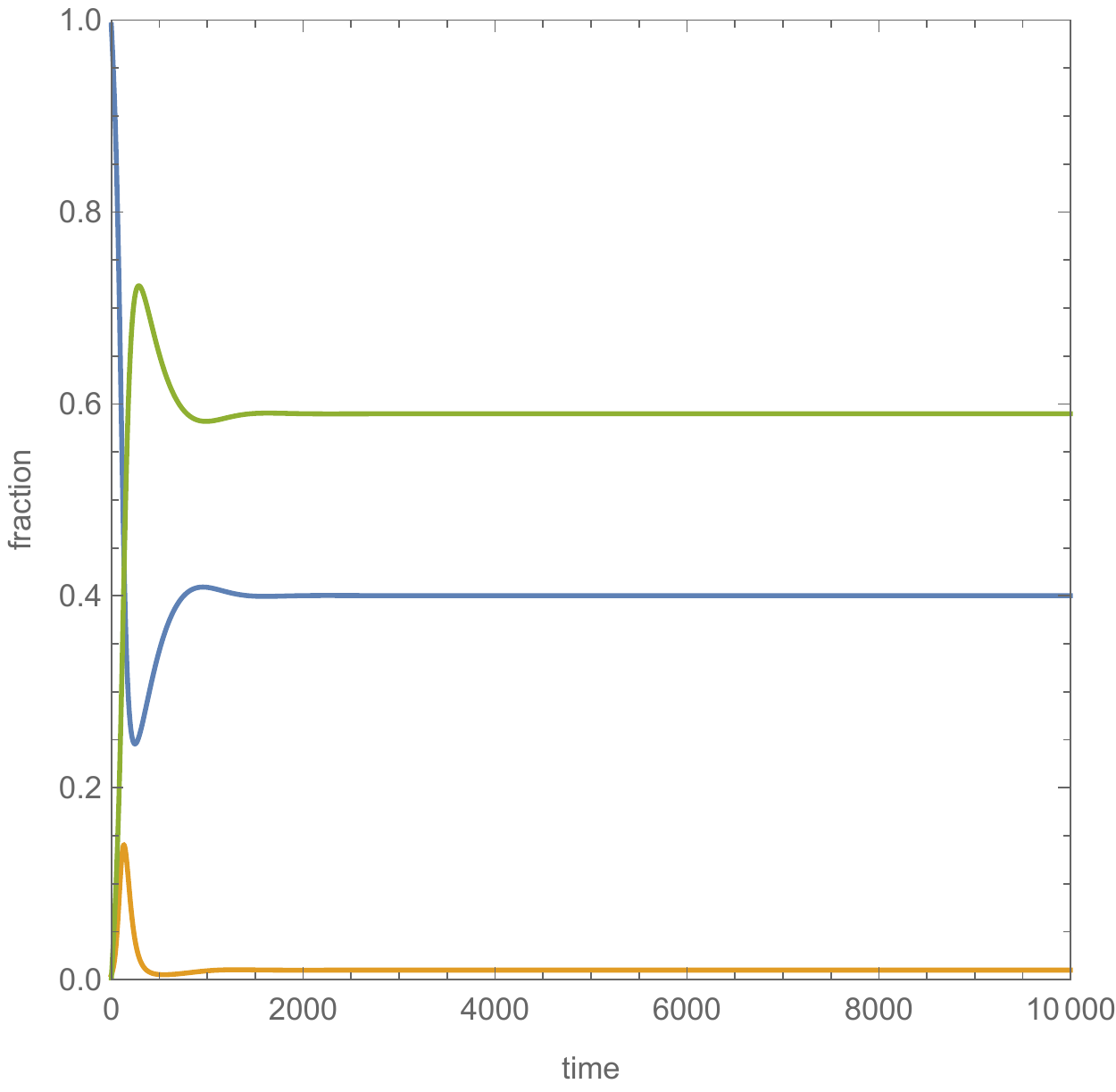}
}
  \put(150,500){
  \includegraphics[width=0.27\textwidth]{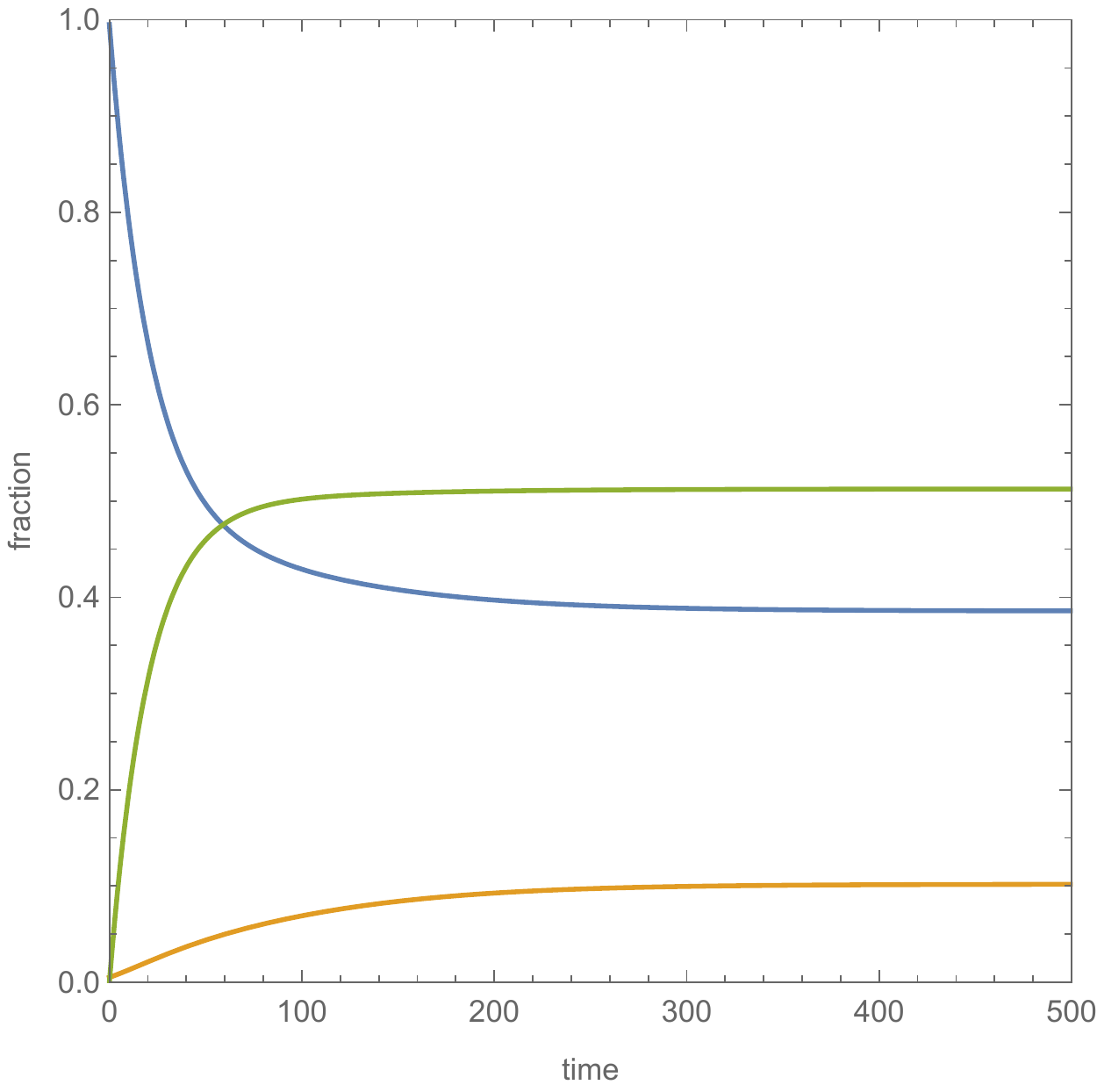}
}
\put(320,500){
  \includegraphics[width=0.27\textwidth]{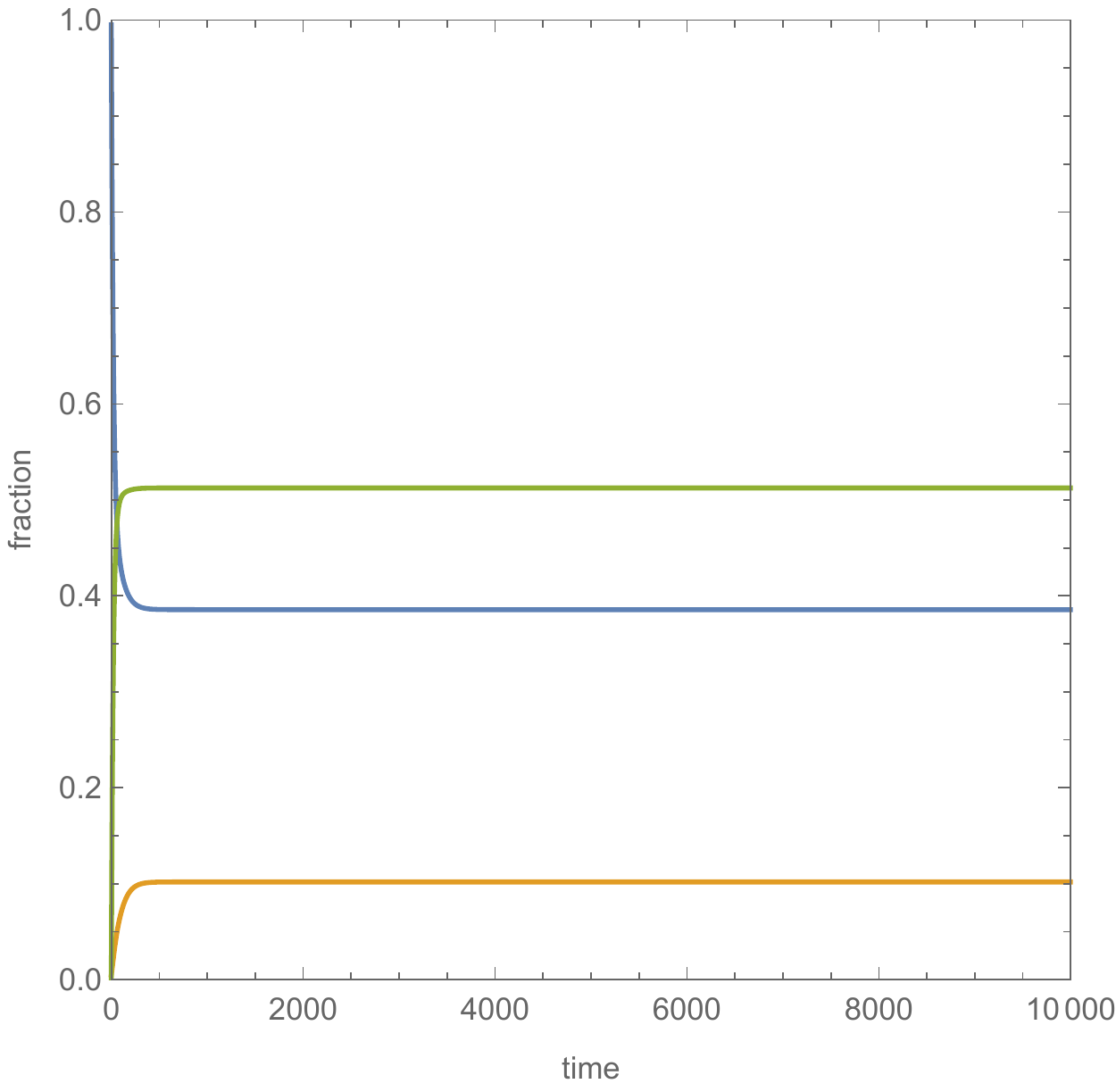}
}
\put(-10,15){
e.
}
\put(-10,145){
d.
}
\put(-10,275){
c.
}
\put(-10,405){
b.
}

\put(-10,535){
a.
}
\end{picture}
\caption{Evolution of $S$, $I$, $R$ for
 $\b=\frac{3}{40}$, $\g=\frac{1}{30}$, $\a=0.01$.
 Rows: a. $\g_1=\frac{1}{20}$; b. $\g_1=\frac{1}{300}$; c. $\g_1=\frac{1}{1000}$; d. $\g_1=\frac{1}{5000}$; e. $\g_1=\frac{1}{10000}$.
\\Left column: solutions to free SIR \eqref{eq:SIR'} for $t\in[0,500]$;
\\Middle column: solutions to \eqref{eq:SIR_BoltzmannDA} for $t\in[0,500]$;
\\Right column: solutions to \eqref{eq:SIR_BoltzmannDA} for $t\in[0,10000]$.
}
\label{f:009}
\end{figure}

\newpage
We conclude with a remark on the solution to \eqref{eq:SIR:jet}. 
The asymptotic fraction $\bar{S}$ is very weakly dependent on $\g_1$, provided that $\g_1$ has the same order of magnitude  of $\b\cdot m$ and $\g$, or smaller.
Indeed, changing  $\g_1$ influences how long the solution stays close to the solution to \eqref{eq:SIR'}, and it changes the asymptotic fractions $\bar{I}$ and $\bar{R}$ that sum $1-\bar{S}$, but it does not perturb significantly $\bar{S}$.
We report on this in  Table~\ref{t:001} and Fig.~\ref{f:009}. 

\begin{table}[h!]
\centering
\begin{tabular}{c|c|c|c}
\hline\hline
$(\g_1)^{-1}$ & $\bar{S}$ & $\bar{I}$ & $\bar{R}$ \\
\hline\hline
10	&	0.373059	&0.157384	&0.469556
 \\
\hline
20		&0.385659&	{0.101855}&	{0.512484}
 \\
 \hline
50	&	0.394638&	0.0508194	&0.554544
 \\
 \hline
100	&	0.397964&	0.0278859&	0.574149
 \\
 \hline
300	&	0.400297&	0.00996697	&0.589736
 \\
 \hline
1000	&	0.401138&	0.00306936&	0.595793
 \\ 
 \hline
 5000	&	0.401429&	0.000619549	&0.597951 \\ 
 \hline
10000	&	0.401466&	0.000310134	&0.598224\\
\hline\hline
\end{tabular}
\caption{Asymptotic values of $S$, $I$, $R$ as  a function of $\g_1$, Eq.\,\eqref{eq:SIR:jet}
with 
 $\b=\frac{3}{40}$, $\g=\frac{1}{30}$, $\a=0.01$.
}
\label{t:001}
\end{table}

%
%
%
%
%
%
%
%
%
%

\newpage

%

%
%
%
%
%
%
%
%
%
%
%

\end{document}